\documentstyle[12pt,a41,epsfig,portland,rotating]{article}

\newcommand{\Mvec}{\mbox{\rm\bf M}}
\newcommand{\Nvec}{\mbox{\rm\bf N}}

\newcommand{\beq}{\begin{equation}}
\newcommand{\eeq}{\end{equation}}
\newcommand{\bea}{\begin{eqnarray}}
\newcommand{\eea}{\end{eqnarray}}

\newcommand{\lsim}{\raisebox{-0.07cm}{$\, \stackrel{<}{{\scriptstyle
\sim}}\, $}}

\newcommand{\EM}{{\rm E-}}

\newcommand\MeV{\,\mbox{MeV}}
\newcommand\GeV{\,\mbox{GeV}}

\newcommand\ONE{\mbox{\boldmath $1$}}
\newcommand\evec{\mbox{\boldmath $e$}}
\newcommand\Lvec{\mbox{\boldmath $L$}}
\newcommand\Rvec{\mbox{\boldmath $R$}}

\newcommand\Pvec{\mbox{\boldmath $P$}}
\newcommand\Uvec{\mbox{\boldmath $U$}}

\newcommand{\Gqqa}{\gamma_{qq}^{N(0)}}
\newcommand{\Gqga}{\gamma_{qg}^{N(0)}}
\newcommand{\Ggqa}{\gamma_{gq}^{N(0)}}
\newcommand{\Ggga}{\gamma_{gg}^{N(0)}}
\newcommand{\Gqqb}{\gamma_{qq}^{N(1)}}
\newcommand{\Gqgb}{\gamma_{qg}^{N(1)}}
\newcommand{\Ggqb}{\gamma_{gq}^{N(1)}}
\newcommand{\Gggb}{\gamma_{gg}^{N(1)}}

\newcommand{\Cq}{C_{2,q}^{N(1)}}
\newcommand{\Cg}{C_{2,g}^{N(1)}}
\newcommand{\bb}{{\beta_1 \over \beta_0}}

\newcounter{lin}

\begin{document}
\begin{titlepage}

\begin{flushleft}
DESY 01--087 \hfill {\tt hep-ph/0203155} \\
February  2002 \\
\end{flushleft}
\sloppy

\vspace{3cm}
\noindent
\begin{center}
{\LARGE\bf QCD Analysis of Polarized Deep Inelastic}\\

\vspace*{2mm}
\noindent
{\LARGE\bf Scattering Data and
Parton Distributions }
\end{center}
\begin{center}

\vspace{4cm}
{\large Johannes Bl\"umlein and Helmut B\"ottcher}

\vspace{2cm}
{\it 
Deutsches Elektronen Synchrotron, DESY}\\

\vspace{3mm}
{\it  Platanenallee 6, D--15738 Zeuthen, Germany}\\

\vspace{2cm}
\end{center}
\begin{abstract}
\noindent
A QCD analysis of the world data on polarized deep inelastic scattering 
is presented in leading and next-to-leading order. New parameterizations 
are derived for the quark and gluon distributions for the kinematic range
$x~\epsilon~[10^{-9},1],~Q^2~\epsilon~[1,10^6]~\GeV^2$. The extrapolation
far outside the domain of the current measurements is given both to allow
for applications at higher values of $Q^2$ and to be able to calculate
integral properties of the present distributions.
The values of
$\Lambda_{\rm QCD}$ and $\alpha_s(M_z)$ are determined. Emphasis is put on
the derivation of the fully correlated $1\sigma$ error bands for these
distributions, which are also given in terms of parameterizations and are
directly applicable to determine experimental errors of other polarized 
observables. The impact of the variation of both the renormalization and 
factorization scales on  the value of $\alpha_s$ is studied. Finally we
perform a factorization--scheme invariant QCD analysis based on the 
observables $g_1(x,Q^2)$ and $d g_1(x,Q^2)/d \log(Q^2)$ in 
next-to-leading order, which is compared to the standard analysis. A 
series of low moments of parton densities, accounting for error 
correlation, are given to allow for comparison with results from lattice
simulations.
\end{abstract}

\end{titlepage}

\newpage
\sloppy

\section{Introduction}
\label{sec:intro}

\vspace*{1mm}
\noindent
The nature of the short-distance structure of polarized nucleons is one
of the central questions of present day hadron physics. Nucleons are
composite fermions and their spin should be obtained as a superposition
of the spins and angular momenta of its constituents, the quarks and 
gluons. At large enough space-like 4-momentum transfers $-q^2 = Q^2$ and 
large energy transfer $\nu = 2 p.q/M$, with $p$ and $M$ the nucleon 
4-momentum and mass, the QCD--improved parton model applies. A 
QCD analysis at leading twist level allows to determine the polarized
parton densities and the QCD scale     $\Lambda_{\rm QCD}$ up to
next-to-leading order (NLO)~[1--3].
The distribution functions
derived can be used to calculate other leading twist hard scattering 
cross sections of polarized nucleons.

During the recent years the remarkable growth of deep inelastic scattering
data off polarized targets~[4--13]
allows to perform 
a detailed analysis. Here the different systematic effects of the data
have to be taken into account. Previous analyzes
[14--16]~\footnote{%
In \cite{LEAD} another factorization scheme is referred to.
For a list of older parameterizations see~\cite{OLDER}.} present
parameterizations only for the central values of the parton distributions.
It is, however, desirable to derive at the same time parameterizations 
of the polarized parton densities and those of their $1\sigma$ error, 
taking into account all error correlations being available. With these
parameterizations at hand it becomes possible to give estimates on the
uncertainty of the measurement of other hard scattering processes of
polarized targets w.r.t. the knowledge of the parton densities. In the 
analysis the polarized parton distributions are first determined at a 
reference scale $Q_0^2$. The aforementioned parameterizations are found 
evolving both the central values and their
errors from this scale to all scales
$Q^2$ of interest, which we choose as $1 \GeV^2  \leq Q^2 \leq 10^6 
\GeV^2$ for $x = Q^2/2p.q~\epsilon~[10^{-9},1]$, and are given in form of
fast and accurate grid interpolations.

Unlike the case in most of the
previous analyzes the QCD  analysis performed in the
present paper determines the QCD--scale $\Lambda_{\rm QCD}$ along with the
parameters of the parton densities at the initial scale $Q_0^2$. Moreover,
positivity constraints are not imposed from the beginning but first left
to the fit. A systematic investigation of the $\chi^2$--profiles during
the analysis showed, which of the parameters of the parton densities
have enough sensitivity to be measured from the current polarized
deep-inelastic scattering world data. In this context it turns out that
the current inclusive data do not yet allow to pin down the flavor
structure of the sea quarks. Therefore we are going to use flavor
$SU(3)$ in the current analysis and leave a refinement to a further
analysis including as well semi--inclusive data.\footnote{The latter
analysis can only be performed if the relevant            fragmentation
functions are known, cf.~\cite{GR}.}
As the subsequent
analysis shows, the small-$x$ behavior of the parton densities cannot
be reliably fixed using the current data. Special dynamical 
effects~\cite{SX,BV1}
may be present at smaller values of $x \lsim 5 \cdot 10^{-3}$ than
currently probed, however, reliable estimates cannot be given at present
due to the lack of calculations on higher order resummed corrections
in this kinematic range. Therefore we present two sets of parameterizations
which could be both accommodated to the present data.

The standard analysis is performed using the $\overline{\rm MS}$
factorization and renormalization scheme. In this way two independent
scales, $\mu^2_f$ and $\mu_R^2$, are introduced into the parton densities.
The observables, $g_1^p(x,Q^2)$ and $g_1^n(x,Q^2)$, are independent of
these scales which leads to two associated (matrix--valued) 
renormalization group equations for the Wilson--coefficients and parton
densities,  which are the evolution equations. Referring to
also other observables, as e.g. the slopes $\partial g_1^{p,n}(x,Q^2)/
\partial \log Q^2$, it is possible to each order in the coupling constant
$a_s =\alpha_s/(4\pi)$ to formulate {\sf physical} evolution equations
\cite{FP,DLY} of observables itself, which removes the dependence on
$\mu^2_f$. This method of analysis is conceptually of high interest
in future analyzes, having precise measurements of the respective
observables at the scale $Q^2_0$ at hand. Then the {\it only} parameter 
to be fitted is $\Lambda_{\rm QCD}$. We also perform this 
factorization--scheme independent analysis in next-to-leading order to 
check the stability of the fitted values of $\Lambda_{\rm QCD}$.

Currently the measurement of also the moments of polarized
parton densities on the lattice become more precise~[23--25].
Yet a series
of systematic and algorithmic effects has to be studied in detail, but 
one may expect that direct comparison between the moments of parton 
densities being extracted out of the world  data and the lattice results
will be possible in the future. This comparison has to include the
proper treatment of errors both on the side of the lattice calculations
and the perturbative analysis.
This usually includes correlated propagation of the experimental errors
through the evolution equations. To allow for comparisons with results 
from lattice calculations we provide a set of lowest moments of parton
densities and their combinations including their errors in the present
analysis.

The paper is organized as follows. In sections~2 and 3 the standard and
the factorization--scheme independent analyzes to next-to-leading order
are described. A summary on the experimental data used in the present 
analysis is given in section~4. Section~5 contains the details on the 
parameterizations of the parton densities and that of their 
$1\sigma$  errors. In Section~6
the results of the standard and factorization--scheme
independent analysis on the parton densities and their errors at the 
scale $Q_0^2$ as well as their evolution throughout the kinematic
range of the parameterization $1 \leq Q^2 \leq 10^6 \GeV^2$ are presented
and the respective values of $\Lambda_{\rm QCD}$ and $\alpha_s(M_Z^2)$
are given and are discussed w.r.t the choice of the renormalization
and factorization scales. The covariance matrices of the parameters
determined in the QCD  fits are provided.  In section~7 a series
of values of the lower moments of the parton densities and their errors, 
suitable
for comparisons with results of lattice simulations, are calculated.
Section~8 contains the conclusions. An appendix provides informations
on the numerical parameterization of different sets of polarized parton
densities and their errors.
\section{Standard Analysis}
\label{sec:sta}

\vspace{2mm}
\noindent
The twist--2 contributions to the structure function $g_1(x,Q^2)$ can
be represented in terms of a {\sc Mellin} convolution of the polarized 
parton densities $\Delta f_i$ and the coefficient functions $\Delta C_i^A$ 
by
\begin{eqnarray}
\label{eqg1}
g_1(x,Q^2)&=& \frac{1}{2}   \sum_{j=1}^{N_f} e_j^2 \int_{x}^1
{dz \over z} \Bigg [ { 1 \over N_f}\, \Delta \Sigma
\left({x \over z},\mu_f^2\right) \Delta C_q^S
\left(z,{Q^2 \over \mu_f^2}\right) 
 + \Delta G\left({x \over z},\mu_f^2\right)
\nonumber\\[2ex]
&&~~~~~~~~~~
\times \Delta C^G\left(z,{Q^2 \over \mu_f^2}\right)
+ \Delta   {q_j}^{NS}\left({x \over z},\mu_f^2\right)
\Delta C_q^{NS}\left(z,{Q^2 \over \mu_f^2}\right) \Bigg ]~,
\end{eqnarray}
where $e_j$ denotes the charge of the $j$th quark flavor and $N_f$ 
is the number of flavors. The scale $\mu_f$ denotes the
factorization scale which is introduced to remove the collinear
singularities from the partonic structure functions. In addition the
above quantities are dependent on the renormalization scale $\mu_r$ of
the strong coupling constant $a_s(\mu_r^2) = g_s^2(\mu_r^2)/(16 \pi^2)$.
The structure function $g_1(x,Q^2)$, as a physical observable, is
independent of the choice of both scales. The parton densities and the
coefficient functions are dependent on these scales and obey corresponding
renormalization group equations.

The singlet and non--singlet parton densities which occur in 
Eq.~(\ref{eqg1}) are expressed by the individual flavor contributions as
\begin{eqnarray}
\label{eqfS}
\Delta \Sigma\left(z,\mu_f^2\right)&=&\sum_{i=1}^{N_f}
\Bigg[\Delta f_{q_i}\left(z,\mu_f^2\right)+
\Delta f_{\bar q_i}\left(z,\mu_f^2\right)\Bigg]\,,
\end{eqnarray}
and
\begin{eqnarray}
\label{eqfNS}
\Delta   {q_i}^{NS}\left(z,\mu_f^2\right)&=&
\Delta f_{q_i}\left(z,\mu_f^2\right)+
\Delta f_{\bar q_i}\left(z,\mu_f^2\right)
-{1 \over N_f} \Delta \Sigma\left(z,\mu_f^2\right)\,,
\end{eqnarray}
respectively, where $f_{q_i}$ denotes the polarized quark distribution
of the $i$th flavor.

The running coupling constant is obtained as the solution of 
\begin{eqnarray}
\label{eqarun}
\frac{d a_s(\mu_r^2)}{d \log(\mu_r^2)}  = - \beta_0 a_s^2(\mu_r^2)
- \beta_1 a_s^3( \mu_r^2) + O(a_s^4)~,
\end{eqnarray}
where the coefficients of the $\beta$--function are given by
\begin{eqnarray}
\label{eqbeta}
\beta_0 &=&  \frac{11}{3} C_A - \frac{4}{3} T_F N_f~, \nonumber\\
\beta_1 &=&  \frac{34}{3} C_A^2 - \frac{20}{3} C_A T_F N_f -
4 C_F T_F N_f~,
\end{eqnarray}
in the $\overline{\rm MS}$--scheme. Here $C_A = 3, T_F = 1/2, C_F =4/3$
and $N_f$ denotes the number of active flavors. In the QCD  fit to the
data we extract $\Lambda_{\rm QCD}^{(4)}$ and choose $N_f = 4$ in 
Eq.~(\ref{eqbeta}) whereas only the three light flavors are used to 
represent $g_1(x,Q^2)$,~Eq.~(\ref{eqg1}). The expression for
$\Lambda_{\rm QCD}^{\overline{\rm MS}}$ is given by
\begin{eqnarray}
\label{eqlam}
\Lambda_{\rm QCD}^{\overline{\rm MS}} 
= \mu_r \exp \left\{-\frac{1}{2} \left[
\frac{1}{\beta_0 a_s(\mu_r^2)} - \frac{\beta_1}{\beta_0^2} \log
\left(\frac{1}{\beta_0 a_s(\mu_r^2)} + \frac{\beta_1}{\beta_0}\right)
\right]\right\}.
\end{eqnarray}
Subsequently we will also compare the values of $a_s$ at the mass scale 
of the $Z$-boson, $a_s(M_Z^2)$. This is obtained matching the values of
$a_s$ at the charm-- and
bottom--quark threshold $M_c = 1.4 \GeV$,
$M_b = 4.5 \GeV$ using Eq.~(\ref{eqlam})
for the value of $\Lambda_{\rm QCD}^{\overline{\rm MS}}$.

The change of the parton densities with respect to the factorization 
scale $\mu_f^2 = Q^2$
is described by  (matrix--valued)  renormalization group equations,
the evolution equations, which read
\begin{eqnarray}
\label{eqNSE}
\frac{\partial \Delta f_{q_i}^{\rm NS}(x,Q^2)}{\partial \log Q^2} &=&
P_{NS}^-(x,a_s) \otimes \Delta f_{q_i}^{\rm NS}(x,Q^2)\;, \\
\label{eqSIE}
\frac{\partial}{\partial \log Q^2} \left(\begin{array}{c} 
\Delta \Sigma(x,Q^2) \\ \Delta G(x,Q^2) \end{array} \right)
&=& \Pvec(x,a_s) \otimes
\left(\begin{array}{c} 
\Delta \Sigma(x,Q^2) \\ \Delta G(x,Q^2) \end{array} \right),
\end{eqnarray}
with
\begin{eqnarray}
\label{eqNSP}
P_{\rm NS}^-(x,a_s) &=& a_s P_{\rm NS}^{(0)}(x)
+ a_s^2 P_{\rm NS}^{-(1)}(x) + {\cal{O}}(a_s^3)\;, \\
\label{eqSIP}
\Pvec(x,a_s) &\equiv&  \left(\begin{array}{cc}
P_{qq}(x,Q^2) & P_{qg}(x,Q^2) \\ P_{gq}(x,Q^2) & P_{gg}(x,Q^2) 
\end{array} \right) = a_s \Pvec^{(0)}(x) + a_s^2 \Pvec^{(1)}(x) +
{\cal{O}}(a_s^3)\;,
\end{eqnarray}
and $\otimes$ the {\sc Mellin} convolution
\begin{eqnarray}
\label{eqmeco}
[A \otimes B](x) = \int_0^1 dx_1 dx_2 \delta(x-x_1 x_2)
A(x_1) B(x_2)\;.
\end{eqnarray}
The polarized coefficient functions and anomalous dimensions were 
calculated in~[1-3]
to next-to-leading order. 

The evolution equations (\ref{eqNSE},\ref{eqSIE}) may be rewritten
changing the evolution variable $\log(Q^2)$ into $a_s$ using
Eq.~(\ref{eqarun}):
\begin{eqnarray}
d \log(Q^2) = -\frac{da_s}{\beta_0 a_s^2+\beta_1 a_s^3}~.
\end{eqnarray}

We solve the evolution equations in {\sc Mellin}--$N$ space. For this
purpose a {\sc Mellin}--transformation
\begin{eqnarray}
\label{eqMEL}
\Mvec[f](N) = \int_0^1 dx x^{N-1} f(x),~~N~\epsilon~\Nvec
\end{eqnarray}
of the equations is carried out under which the {\sc Mellin} convolution
$\otimes$ turns into the ordinary product. After the transformation was
performed the argument $N$ is analytically continued to the complex
plane. The principle way of solution of the Eqs.~(\ref{eqNSE},\ref{eqSIE}) 
is described in the literature in detail, see \cite{FP,GRV,BV1},
which we summarize briefly. For the non--singlet case the solution to
NLO is
given by
\begin{eqnarray}
\label{eqNSS}
\Delta q^{\rm NS}_i(N,a_s) = \left(\frac{a_s}{a_0}\right)
^{-P_{\rm NS}^{(0)}/\beta_0} \left[1 - \frac{1}{\beta_0}(a_s-a_0)
\left(P_{\rm NS}^{-(1)} - \frac{\beta_1}{\beta_0} P_{\rm NS}^{(0)}\right)
\right] \Delta q^{\rm NS}_i(N,a_0)~,
\end{eqnarray}
where $a_s = a_s(Q^2), a_0 = a_s(Q_0^2)$.  The singlet solution reads
\begin{eqnarray}
\label{eqSIS}
\left(\begin{array}{c} \Delta \Sigma(N,a_s) \\ \Delta G(N,a_s)
\end{array} \right)
= \left[ \ONE + a_s \Uvec_1(N)\right] \Lvec(N,a_s,a_0) \left[\ONE -
a_0 \Uvec_1(N)\right]
\left(\begin{array}{c} \Delta \Sigma(N,a_0) \\ \Delta G(N,a_0)
\end{array} \right)~.
\end{eqnarray}
Here the leading order evolution matrix $\Lvec$ is given by
\begin{eqnarray}
\label{eqLOS}
\Lvec(a_s,a_0,N) = \evec_-(N) \left(\frac{a_s}{a_0}\right)^{-r_-(N)}
                 + \evec_+(N) \left(\frac{a_s}{a_0}\right)^{-r_+(N)}~,
\end{eqnarray}
with eigenvalues of the LO singlet evolution matrix
\begin{eqnarray}
\label{eqLOEVA}
r_{\pm} =  \frac{1}{\beta_0} \left [ tr(\Pvec^{(0)}) \pm \sqrt{
tr(\Pvec^{(0)})^2 - det_2(\Pvec^{(0)})}\right]
\end{eqnarray}
and the eigenvectors
\begin{eqnarray}
\label{eqLOEVE}
\evec_{\pm} = \frac{\Pvec^{(0)}/\beta_0 - r_\mp \ONE}{r_\pm - r_\mp}~.
\end{eqnarray}
The       matrix
$\Uvec_1(N)$ is given by
\begin{eqnarray}
\label{eqUNLO}
\Uvec_1(N) = - \evec_- \Rvec_1 \evec_- - \evec_+ \Rvec_1 \evec_+
+ \frac{\evec_+ \Rvec_1 \evec_-}{r_- -r_+ -1} 
+ \frac{\evec_- \Rvec_1 \evec_+}{r_+ -r_- -1}
\end{eqnarray}
with 
\begin{eqnarray}
\Rvec_1 = [\Pvec^{(1)} - (\beta_1/\beta_0) \Pvec^{(0)}]/\beta_0~.
\end{eqnarray}
The input distributions $\Delta q^{\rm NS}_i(N,a_0), \Delta \Sigma(N,a_0)$
and  $\Delta G(N,a_0)$, see section~5, are evolved to the scale
$Q^2$, respectively to the coupling $a_s(Q^2)$. The inverse {\sc
Mellin}--transform to $x$--space is performed by a contour integral in
the complex plane around all singularities, which can be written as
\begin{eqnarray}
\label{invers}
f_{\rm TH}(x) = \frac{1}{\pi} \int_0^{\infty} dz~{\sf Im} \left[
\exp(i\phi) x^{-c(z)} f_{\rm TH}[c(z)]\right ]
\end{eqnarray}
applying symmetry properties of the integrand. In practice an integral 
along the path $c(z) = c_1 + \rho[\cos(\phi) + i \sin(\phi)]$, with 
$c_1 = 1.1, \rho \ge 0$
and $\phi = (3/4) \pi$ can be used. 
The upper bound on $\rho$ is to be chosen by sufficient
numerical convergence of the integral~(\ref{invers}). The theoretical
prediction $f_{\rm TH}(x)$ for the respective observables depends on the
parameters of the parton distributions chosen at the starting scale
$Q_0^2$ and on $\Lambda_{\rm QCD}$. These parameters are determined by
a fit to the data using the $\chi^2$--method, see section~6.
\section{Scheme Invariant Evolution of Polarized Structure Functions}
\label{sec:si}

\vspace{2mm}
\noindent
In the foregoing section we investigated the QCD evolution of the
polarized structure function $g_1(x,Q^2)$ at the level of twist--2
using the conventional picture of the QCD improved parton model. This
requires choices of the non--perturbative partonic input distributions at
a starting scale $Q_0^2$. One outcome of the analysis is that some of
the distributions are still very difficult to determine by a fit to
the polarized
deep inelastic data. This particularly applies to the polarized
gluon and
sea--quark densities. 

The scaling violations of deep inelastic structure functions, i.e.
their $Q^2$--behavior, are properties of observables, and do thus 
{\it not} depend on particular representations, as scheme--dependent
decompositions, which occur applying the parton model. Instead referring
to the latter one might wish to eliminate the scheme dependence as
occurring due to mass factorization all together and describe the
evolution of polarized structure functions based on {\it observables}
only. As a consequence, no parton distributions would even emerge in
this description, and choices of the parameterization of less constraint
input densities can
be avoided.\footnote{After having carried out the analysis projections
onto the various parton--densities, including the gluon density, in
{\it whatever} factorization scheme, the $\overline{\rm MS}$ scheme or
other schemes
\cite{ALTA,LEAD},
 are possible  including the
respective experimental errors.}

Up to NLO the evolution of the structure function $g_1(x,Q^2)$ is
described by one non--singlet~(\ref{eqNSE}) and the coupled singlet
evolution equations~(\ref{eqSIE}). Instead of Eq.~(\ref{eqNSE}) one may
use the scheme--invariant equation based on the non--singlet
structure function     $g_1^{\rm NS}(x,Q^2)$
\begin{eqnarray}
\label{eqNS1}
\frac{\partial g_1^{\rm NS}(x,Q^2)}{\partial t} = K^{\rm NS}(a,x)
  \otimes      g_1^{\rm NS}(x,Q^2)
\end{eqnarray}
with\footnote{Note that the splitting functions and coefficient functions
are expressed in such a way that all evolution equations refer to
logarithmic variations in $Q^2$ and $a(Q^2)$ is used as the coupling
constant.}
\begin{eqnarray}
\label{eqKNS1}
K^{\rm NS}(a,x) = \frac{a}{2}\left\{P^{\rm NS}_0(x) + 
a\left[P_1^{\rm NS}(x)
- \frac{\beta_1}{\beta_0} P_0^{\rm NS}(x) - \beta_0 c_{q,1}(x)\right]
\right\}~.
\end{eqnarray}
Here the evolution variable is chosen as
\begin{equation}
\label{eqT}
t=-{2 \over \beta_0} \ln\left({a_s(Q^2) \over a_s(Q_0^2)}\right)~.
\end{equation}

For the singlet--evolution two observables have to be selected to form
the factorization--scheme invariant evolution equations. In case of
the polarized structure functions a natural choice is to use the 
flavor--singlet part of $g_1(x,Q^2)$ and its partial derivation w.r.t. 
$t$.~\footnote{Evolution equations of this type have been discussed before
in Ref.~\cite{FP}. In the unpolarized case one may consider the pair
$F_2(x,Q^2), F_L(x,Q^2)$ as well, which has been considered in detail in
Ref.~\cite{DLY}. For similar approaches, partly derived only for the
small-$x$ domain, see~\cite{SIO}. Corresponding evolution equations
for time--like virtualities, which describe fragmentation in a 
factorization--scheme invariant way, were given in \cite{DLY}.}
The singlet evolution equation reads (taking the {\sc Mellin} transform)
\begin{equation}
\label{eqevo2}
{\partial \over \partial t}\left( \begin{array}{c}
F^N_A\\
{F^N_B}
\end{array} \right) =
-{1 \over 4}  {\bf K}_S^{g_1}(N)
\left( \begin{array}{c}
F^N_A\\
{F^N_B}
\end{array} \right)  =
-{1 \over 4}
\left( \begin{array} {cc}
K^N_{AA} & K^N_{AB}\\
K^N_{BA} & K^N_{BB}
\end{array} \right)
\left( \begin{array}{c}
F^N_A\\
{F^N_B}
\end{array} \right)~,
\end{equation}
with $F_A = g_1(N,Q^2), F_B = \partial g_1(N,Q^2)/\partial t$.
The {\sf physical} evolution kernels are given in leading order by
\begin{equation}
\label{eqG21}
\begin{array}{rclcrcl}
K_{22}^{N(0)}&=&0~, & \hspace{.5cm}&
K_{2d}^{N(0)}&=&-4~, \nonumber \nonumber \\ \\
K_{d2}^{N(0)}&=&\displaystyle{{1 \over 4} \Bigg(\Gqqa \Ggga-\Gqga 
\Ggqa\Bigg)}~, & \hspace{.5cm}&
K_{dd}^{N(0)}&=&\displaystyle{\Gqqa+\Ggga }~.\nonumber
\nonumber 
\end{array}
\end{equation}
In next-to-leading order they read
\begin{eqnarray}
\label{eqG22}
K_{22}^{N(1)}&=&0~,
\\[2ex]
K_{2d}^{N(1)}&=&0~,
\\[2ex]
K_{d2}^{N(1)}&=&{1 \over 4} \Bigg[\Ggga \Gqqb+\Gggb \Gqqa -\Gqgb 
\Ggqa -\Gqga \Ggqb\Bigg]
\nonumber\\[2ex]
&& -{\beta_1 \over 2 \beta_0} \Bigg(\Gqqa \Ggga-\Ggqa \Gqga\Bigg)
   +{\beta_0 \over 2} \Cq \Bigg(\Gqqa+\Ggga-2 \beta_0\Bigg)
\nonumber\\[2ex]
&&-{\beta_0 \over 2} {\Cg \over \Gqga} \Bigg[(\Gqqa)^2-\Gqqa\Ggga+2 
\Gqga\Ggqa-2 \beta_0 \Gqqa\Bigg]
\nonumber\\[2ex]
&&-{\beta_0 \over 2} \Bigg(\Gqqb-{\Gqqa \Gqgb \over \Gqga} \Bigg)~,
\\[2ex]
\label{eqG23}
K_{dd}^{N(1)}&=&\Gqqb+\Gggb-\bb \Bigg(\Gqqa+\Ggga \Bigg)
\nonumber\\[2ex]
&& -{2 \beta_0 \over \Gqga} \Bigg[ \Cg \Big(\Gqqa - \Ggga-2 
\beta_0\Big)-\Gqgb\Bigg]
  +4 \beta_0 \Cq -2 \beta_1~.
\end{eqnarray}
Here $\gamma_{ij}^{N(0,1)}$ denotes the polarized deep--inelastic
anomalous dimensions
\begin{eqnarray}
\gamma_{ij}(a,N) &=& \sum_{k=0}^\infty a^{k+1} \gamma_{i,j}^{(k)}(N)~,\\
\gamma_{ij}^{N(0,1)} &=& - 2 \int_0^1 dz z^{N-1} P_{ij}^{(0,1)}(z)~,
\end{eqnarray}
and $C_{2,q(g)}^{N(1)}$ is the {\sc Mellin}--transform of the Wilson 
coefficients
\begin{eqnarray}
C_{2,k}^{N(1)} &=& \int_0^1 dz z^{N-1} C_{2,k}^{(1)}(z)~.
\end{eqnarray}
For this combination in next-to-leading order the evolution depends on 
two evolution kernels only.

The kernels $K^{\rm NS}$ and ${\bf K}_{\rm S}^{g_1}$ are factorization
scheme--invariant quantities. Their analytic structure in $z$--space
is difficult to obtain due to the inverse {\sc Mellin}--convolutions
of inverse coefficient functions being required. Already in the case of
the inverse of the leading order splitting function $P_{gq}(z)$
a rather complicated expression is found~\cite{BK}. Therefore we
perform all computations in {\sc Mellin}--$N$ space, where the
physical evolution kernels are polynomials out of anomalous dimensions and
coefficient functions. Their analytic continuation to complex values
of $N$ can be performed using the respective representations of harmonic
sums at high numerical precision, cf.~\cite{MEL}. In the unpolarized case
approximate representations were also obtained in \cite{NV}.

The advantage of studying factorization--scheme invariant evolution 
equations both in the non--singlet and the singlet case is that the
input distributions are observables. Although the present world 
statistics is too low, future high statistics measurements  may provide 
accurate input densities $g_1(x,Q^2_0)$ and $dg_1(x,Q^2_0)/dt$. In this
case the {\sf only} parameter to be measured analyzing the scaling
violations of $g_1(x,Q^2)$ is
$\Lambda_{\rm QCD}$. At present the solution of the physical evolution
equations in the polarized case cannot yet take full advantage of the
method, since the respective observables are not yet measured well enough
at typical input scales $Q_0^2$. This different formulation, however,
leads to an alternative view on the data in extracting 
$\Lambda_{\rm QCD}$, as different input densities, $g_1(x,Q^2_0)$ and 
$dg_1(x,Q^2_0)/dt$, are fitted as compared to $\Delta \Sigma(x,Q_0^2)$
and $\Delta G(x,Q_0^2)$ in the standard analysis. A comparison of the
values of $\Lambda_{\rm QCD}$ obtained in both analyzes may indicate
the stability of the determination of the QCD--parameter.
\section{Data}
\label{sec:dat}

\vspace{2mm}
\noindent
The remarkable growth of experimental data on inclusive polarized deep 
inelastic scattering of leptons off nucleons over the last years allows
to perform  refined QCD analyzes of polarized structure functions in 
order to reveal the spin--dependent partonic structure of the nucleon.
For the QCD analysis presented in the present paper the following data 
sets are used:
the EMC proton data \cite{EMCp}, the E142 neutron data \cite{E142n},
the HERMES neutron data \cite{HERMn}, the E154 neutron data \cite{E154n},
the SMC proton and deuteron data \cite{SMCpd}, the E143 proton and 
deuteron data \cite{E143pd}, the HERMES proton data \cite{HERMp}, the 
E155 deuteron data \cite{E155d}, and the E155 proton data 
\cite{E155p}\footnote{Earlier data from Ref.~\cite{E80130} are not 
considered.}. The number of the published
data points above $Q^2 = 1.0~\GeV^2$ for the different data sets are
summarized in Table~1 for both the asymmetry data, i.e. $g_1/F_1$ or
$A_1$, and data on $g_1$ together with the $x-$ and $Q^2$--ranges
for the different experiments.\footnote{All corrections to the data
are assumed to be carried out, including the QED radiative corrections
\cite{RC1,RC2}.}
There are 435 data points for asymmetry data, a number which exceeds
the number of data points for $g_1$ data by  a factor of two. We
therefore are utilizing the asymmetry data which are expected to give
a better statistical accuracy. 

The QCD fits are performed on the polarized structure function 
$g_1(x,Q^2)$ which has to be evaluated from the asymmetry data. 
Experimentally cross section
asymmetries for longitudinally polarized lepton scattering off
longitudinally polarized nucleons are measured
\begin{equation}
A_{||} = \frac 
{\sigma^{\uparrow\uparrow} - \sigma^{\uparrow\downarrow}}
{\sigma^{\uparrow\uparrow} + \sigma^{\uparrow\downarrow}}~.
\end{equation}

\noindent
The arrow--combination $\uparrow\uparrow$($\uparrow\downarrow$) denotes
parallel(anti--parallel) spin orientation of the lepton and the nucleon. 
The ratio of structure functions $g_1/F_1$ and the longitudinal virtual 
photon asymmetry $A_1$ are related to $A_{||}$ by

\begin{eqnarray}
\frac{g_1}{F_1}
& = & \frac{1}{(1 + \gamma^2)} \left[ \frac{A_{||}}{D} + 
(\gamma - \eta) A_2 \right], \\
A_1     & = & \frac{A_{||}}{D} - \eta A_2~,
\end{eqnarray} 

\noindent
and the relation between $g_1/F_1$ and $A_1$ is

\begin{equation}
\frac{g_1}{F_1} 
= \frac{1}{(1 + \gamma^2)} \left[ A_1 + \gamma A_2 \right].
\end{equation}

\noindent
Here $A_2$ is the transverse virtual photon asymmetry and
\begin{eqnarray}
 D &=& \frac{1-(1-y)\epsilon}{1+\epsilon R(x,Q^2)}~,\\
\epsilon &=& \frac{4(1-y)-\gamma^2 y^2}{2y^2+4(1-y)+\gamma^2 y^2}~,\\
 \gamma &=& \frac{2Mx}{\sqrt{Q^2}}~,    \\
 \eta &=& \frac{\epsilon \gamma y}{1-\epsilon(1-y)}~.
\end{eqnarray}
$D$ denotes the virtual photon depolarization factor, $\epsilon,
\gamma$ and $\eta$ are kinematic factors, $y=Q^2/(Sx)$ 
is the Bjorken variable, with $S = M^2 + 2 M E_e$, $M$ the nucleon mass,
$E_e$ the electron energy in the target rest frame,
and $R(x,Q^2) = \sigma_L/\sigma_T$.
For $R(x,Q^2)$ the SLAC parameterization $R_{1990}$ \cite{R1990} is used 
by most of the experiments. At the time of the EMC experiment this 
parameterization was not available yet and $R$ was assumed to be $Q^2$ 
independent. The SMC collaboration adopted a combination of $R_{1990}$ 
(for $x > 0.12$) and a parameterization derived by the NMC collaboration 
\cite{RNMC} (for $x < 0.12$). In the E155 experiment a recent SLAC 
parameterization for $R$ \cite{RE143} was used. The changes in the data  
caused by using the different $R$ parameterizations are not significant and 
stay within the experimental errors\footnote{The EMC proton data, where
the biggest impact is expected, change by a few percent only, see 
Ref.~\cite{AAC}.}.

The magnitude of $A_2$ has been measured by SMC \cite{SMCA2}, 
E154 \cite{EE154}, and E143 \cite{E143pd} and 
found to be small. To a good
approximation its contribution to $g_1/F_1$ and $A_1$ which is being
further suppressed by the kinematic factors $\gamma$ and $\eta$ can be
neglected. Nevertheless, E143 and E154
 have exploited their
measurements, E155 has approximated this contribution by $g_2^{\rm
WW}(x,Q^2)$ through the Wandzura--Wilczek
expression \cite{WW},\footnote{Note that this relation holds also in the
presence of quark--
\cite{BT}
 and target--mass corrections~\cite{BT,PR%
}. Related
integral relations for twist--3 contributions  and structure functions
with electro--weak couplings were derived in Refs.~\cite{BT,BLKO}.
Recently these relations have been found to hold also for diffractive
scattering,~\cite{BR}.} which is calculated from the measured structure
function $g_1(x,Q^2)$ in the approximation that twist--2 contributions
are dominant:
\begin{eqnarray}
 g_2^{\rm WW}(x,Q^2) = -g_1(x,Q^2) + \int_x^1 \frac{dz}{z} g_1(z,Q^2)~.
\end{eqnarray}
HERMES has accounted for the $A_2$ contribution to the proton data by
using a fit based on existing measurements \cite{HERMp}. 
To obtain $g_1(x,Q^2)$ one has to multiply the ratio
$g_1(x,Q^2)/F_1(x,Q^2)$ with the unpolarized structure function
$F_1(x,Q^2)$ which can be calculated from the usually measured
unpolarized structure function $F_2(x,Q^2)$ by
\begin{equation}
F_1(x,Q^2) = \frac {(1+\gamma^2)}{2x(1+R(x,Q^2)}F_2(x,Q^2)~.
\end{equation}

\noindent
For all data sets we are using the SLAC $R_{1990}(x,Q^2)$ 
\cite{R1990} and the NMC $F_2(x,Q^2)$ parameterization \cite{F2NMC} to
perform this calculation.

The data sets used contain both statistical and systematic errors except 
the SMC data set which is given with statistical errors only. It is known 
that the systematic errors are partly correlated which would lead to an 
overestimation of the errors when added in quadrature with the statistical 
ones and hence to a reduction of the $\chi^2$ value in the minimization 
procedure. To treat all data sets on the same footing we decided to use the 
statistical errors only. 
However, we allow for a relative normalization shift between the 
different data sets within the normalization uncertainties quoted 
by the experiments. Thereby we are 
taking into account the main systematic uncertainties coming from the 
measurements of the luminosity and  the beam and target polarization. 
The normalization shift for each data set enters as an additional term in 
the $\chi^2$--expression which then reads

\begin{equation}
\chi^2 = \sum_{i=1}^{n^{exp}} \left [ 
         \frac {(N_i - 1)^2}
               {(\Delta N_i)^2} + 
         \sum_{j=1}^{n^{data}} 
         \frac {(N_i g_{1,j}^{data} - g_{1,j}^{theor})^2}
               {(\Delta g_{1,j}^{data})^2} 
         \right ],
\end{equation}

\noindent
where the sums run over all data sets and in each data set over all data 
points. The minimization of the $\chi^2$ value above to determine the best 
parameterization of the polarized parton distributions is done using the 
program {\tt MINUIT} \cite{MINUIT}. Only fits giving a positive definite
covariance matrix at the end have been accepted in order to be able to 
calculate the fully correlated $1\sigma$ error bands.
%
\section{Parameterizations of the Polarized Parton Distributions and 
their Errors}
\label{sec:par-err}
\subsection{Parameterization of the Parton Densities}
\label{subsec:parcho}

\vspace{2mm}
\noindent
The shape chosen for the parameterization of the polarized parton 
distributions in $x$--space at the input scale of $Q^2_0 = 4.0~\GeV^2$ is
\begin{equation}
\label{param}
x\Delta f_i(x,Q_0^2) = \eta_i A_i x^{a_i} (1 - x)^{b_i}
\left(1 + \gamma_i x + \rho_i x^{\frac{1}{2}}\right)~.
\end{equation}

\noindent
The term $x^{a_i}$ controls the low--$x$ behavior of the parton densities
and $(1 - x)^{b_i}$ that at large values of $x$. The remainder polynomial
factor accounts for the additional medium--$x$ degrees of freedom.
The normalization constants $A_i$
\begin{equation}
\label{anorm}
A_i^{-1} = \left( 1 + \gamma_i\frac{a_i}{a_i+b_i+1} \right)
               B(a_i,b_i+1) + \rho_i B\left(a_i+\frac{1}{2},b_i+1\right)
\end{equation}

\noindent
are
chosen such that the
$\eta_i$ are the first moments of $\Delta q_i(x,Q_0^2)$,
$\eta_i =\int_0^1dx \Delta q_i(x,Q_0^2)$. Here $B(a,b)$ is the 
Euler Beta--function being related to the $\Gamma$--function by

\begin{equation}
B(a,b) = \frac {\Gamma(a) \Gamma(b)} {\Gamma(a+b)}~.
\end{equation}
In the present approach the QCD--evolution equations are solved in
{\sc Mellin}--$N$ space as described in section~\ref{sec:sta}.
The {\sc Mellin}--transform of the parton densities is performed and
{\sc Mellin}--N moments are calculated for complex arguments $N$~:

\begin{eqnarray}
\label{MNmom}
\Mvec[ \Delta f(x,Q_0^2)](N) & = & \int_0^1 x^{N-1} \Delta f(x,Q^2_0) dx
\nonumber\\ &=&
\eta_i A_i 
                     \left( 1 + \gamma_i\frac{N-1+a_i}{N+a_i+b_i} \right)
                     B(N-1+a_i,b_i+1)  \nonumber \\ 
                   &   & \nonumber \\
           &   & + \rho_i B\left(N+a_i-\frac{1}{2},b_i+1\right)~.
\end{eqnarray}

\noindent
As seen from Eq.~(\ref{param}) there are five parameters for each parton
distribution. To meet both the quality of the present data and the 
reliability of the fitting program the number of parameters has to be 
reduced. 
Assuming $SU(3)$ flavor symmetry and  a flavor symmetric sea one only
has to derive one general polarized sea--quark distribution. The
first moments of the polarized valence distributions $\Delta u_v$ and 
$\Delta d_v$ can be fixed by the $SU(3)$ parameters $F$ and $D$ as 
measured
in neutron and hyperon $\beta$--decays according to the relations~:

\vspace*{-0.75cm}
\begin{eqnarray}
\eta_{u_v} - \eta_{d_v} & = & F + D~,           \\
\eta_{u_v} + \eta_{d_v} & = & 3F - D~.
\end{eqnarray}

\noindent
A re-evaluation of $F$ and $D$ was performed in Ref.~\cite{AAC} on the
basis of updated $\beta$--decay constants \cite{PDGOLD} leading to
\begin{eqnarray}
\eta_{u_v} &=& ~~0.926 \pm 0.014~, \\
\eta_{d_v} &=& -0.341 \pm 0.018~.
\end{eqnarray}
Given 
the present accuracy of the data a number of parameters is set to zero, 
namely $\rho_{u_v} = \rho_{d_v} = 0$, $\gamma_{\bar q} = 
\rho_{\bar q} = 0$, and $\gamma_G = \rho_G = 0$. This choice reduces the 
number of parameters to be fitted for each polarized parton density to 
three. In addition the QCD parameter $\Lambda_{\rm QCD}$ is to be
determined. 
We allow for a relative normalization shift between the different data
sets within the normalization uncertainties quoted by the
experiments. These normalization shifts were fitted once and then
fixed afterwards. 

The polarized parton densities to be determined are chosen to be 
\begin{eqnarray}
\Delta u_v(x) &=& \Delta u(x) - \Delta \overline{u}(x)~, \nonumber\\
\Delta d_v(x) &=& \Delta d(x) - \Delta \overline{d}(x)~, \nonumber\\
\frac{1}{6}\Delta \overline{Q}(x) &=& \Delta \overline{q}(x) =
\Delta \overline{u}(x) =
\Delta \overline{d}(x) = \Delta {s}(x)
 = \Delta \overline{s}(x)~, \nonumber\\ 
{\rm and}~\Delta G(x)~.
\end{eqnarray}
Twelve parameters representing the parton densities are determined in the
fit. These are~: for $\Delta u_v$: $a_{u_v}$, $b_{u_v}$, and 
$\gamma_{u_v}$,
for $\Delta d_v$: $a_{d_v}$, $b_{d_v}$, and $\gamma_{d_v}$, for 
$\Delta \overline{Q}$: $\eta_{\bar q}$, $a_{\bar q}$, and $b_{\bar q}$, 
and
for $\Delta G$: $\eta_G$, $a_G$, and $b_G$. Note that $\eta_{\bar q}$
represents the first moment of the total quark sea. Starting off with 
these parameters the analysis shows, however, that the four parameters 
$\gamma_{u_v}$, $\gamma_{d_v}$, $b_{\bar q}$, and $b_G$ turn out to 
have very large errors at $\chi^2_{\rm min}$. Altering them within the
error range does not lead to a significant change of $\chi^2$, which
shows that they are badly constrained by the inclusive data used in the
analysis. We, therefore, fixed these parameters at their values obtained
at the end of the fit and consider them as outer model-parameters. 
It is not expected that the small--$x$ behavior of the polarized gluon
and the sea--quarks is much different. To achieve this
it turned out that the small--$x$ slopes of the gluon and the sea--quarks
are to be related like $a_G = a_{\bar q} + C$, with $C = 0.5~...~1.0$.
We therefore decided to fix one of the parameter relative to the other.
In fixing the high--$x$ slopes $b_G$ and $b_{\bar q}$ we adopted a
relation as derived from the unpolarized parton densities, namely
$b_{\bar q} / b_G {\rm (pol)} = b_{\bar q} / b_G {\rm (unpol)}$ (see
e.g. Ref. \cite{GRV}). Both relations together are suited to lead to
positivity for $\Delta G$ and $\Delta {\bar q}$.  
\footnote%
{For a discussion of positivity of parton densities see \cite%
{POSI}.}
No positivity constraint was assumed for $\Delta u_v$ and $\Delta d_v$.
After this the statistical measurement is only applied to the remaining
seven parameters which describe the parton densities and to
$\Lambda_{\rm QCD}$. 
\subsection{Error Calculation}
\label{subsec:errcal}

\vspace{2mm}
\noindent
The evolved polarized parton densities and structure functions are linear
functions of the input densities. 
Let $f(x,Q^2;a_i|_{i=1}^k)$ be the 
evolved density at $Q^2$ depending on the parameters $a_i|_{i=1}^k$. Then 
its correlated error as given by Gaussian error propagation is
\begin{equation}
\label{error}
\Delta f(x,Q^2) = \left[ \sum_{i=1}^k 
                \left( \frac{\partial f}{\partial a_i} \right)^2 C(a_i,a_i) 
                + \sum_{i\neq j=1}^k 
                \left( \frac{\partial f}{\partial a_i}
                \frac{\partial f}{\partial a_j} \right) C(a_i,a_j) 
                \right]^{\frac{1}{2}},
\end{equation}

\noindent
where $C(a_i,a_j)$ are the elements of the covariance matrix determined
in the QCD  analysis at the scale $Q_0^2$. The gradients
$\partial f/\partial a_i$ at this scale can be calculated analytically,
except for $\Lambda_{\rm QCD}$.
Their value at $Q^2$ is calculated by evolution. The general form of the 
derivative of the {\sc Mellin} moment $\Mvec [f(a)](N)$ w.r.t.
the parameter $a$ is 

\begin{equation}
\frac {\partial \Mvec [f(a)](N)}{\partial a} = F(a) 
\times \Mvec [f(a)](N)
\end{equation}

\noindent
for complex values of $N$. Here we give the relevant expressions for 
$F(a)$ as the normalized gradients w.r.t. the parameters being varied in 
the final fit, which are finally two parameters per distribution. For the
polarized parton distributions $\Delta u_v$ and $\Delta d_v$ one obtains

\begin{eqnarray}
\label{grad_N_1}
F(a_i)    & = & \psi(N-1+a_i) - \psi(N+a_i+b_i) \nonumber \\ 
            &   & + \left( \frac{\gamma_i(b_i+1)}
                    {(N+a_i+b_i)(N+a_i+b_i+\gamma_i(N-1+a_i))} \right) 
\nonumber \\
            &   & -\; \psi(a_i) + \psi(a_i+b_i+1) - 
                  \left( \frac{\gamma_i(b_i+1)}
                         {(a_i+b_i+1)(a_i+b_i+1+\gamma_ia_i)} \right) \;,
          \\
F(b_i)    & = & \psi(b_i+1) - \psi(N+a_i+b_i) \nonumber \\
            &   & - \left( \frac{\gamma_i(N-1+a_i)}
                    {(N+a_i+b_i)(N+a_i+b_i+\gamma_i(N-1+a_i))} \right) 
\nonumber \\
            &   & -\; \psi(b_i+1) + \psi(a_i+b_i+1) + 
                  \left( \frac{\gamma_ia_i}
                         {(a_i+b_i+1)(a_i+b_i+1+\gamma_ia_i)} \right) \;,
\end{eqnarray}

\noindent
whereas for $\Delta {\bar q}$ and $\Delta G$

\vspace*{-0.5cm}
\begin{eqnarray}
F(\eta_i) & = & \frac{1}{\eta_i} \;,           \\
\label{grad_N_2}
F(a_i)    & = & \psi(N-1+a_i) - \psi(N+a_i+b_i) - \psi(a_i) + 
                  \psi(a_i+b_i+1)~.
\end{eqnarray}

\noindent
Here $\psi(z) = d/dz(\log\Gamma(z))$ denotes the {\sc Euler} 
$\psi$-function. The gradients evolved in {\sc Mellin} space are then
transformed back to $x$ space and are used according to Eq.~(\ref{error}).
To obtain the gradients for the error calculation of the polarized 
structure functions $g_1^p$ and $g_1^n$, which are composed of the
polarized parton densities the expressions above, 
Eqs.~(\ref{grad_N_1}--\ref{grad_N_2}) 
have to be multiplied with the corresponding Wilson
coefficient functions.

This yields the errors as far as the QCD parameter $\Lambda_{\rm QCD}$ is
fixed and regarded as uncorrelated. The inclusion of the QCD parameter
is performed easiest by numerical methods due to non--linear and 
iterative aspects in the calculation of $\alpha_s(Q^2,\Lambda_{\rm QCD})$.
The respective gradients are calculated performing the evolution
both for $\Lambda \pm \delta$, with $\delta \ll \Lambda$

\begin{equation}
\frac {\partial f(x,Q^2,\Lambda)}{\partial \Lambda} = 
       \frac {f(x,Q^2,\Lambda + \delta)-f(x,Q^2,\Lambda -
\delta)}{2\delta}  
\end{equation}

\noindent
using  values for $\delta \sim  10~\MeV$ in the present analysis.

Finally we also present the gradients at the input scale $Q_0^2$ in
$x$-space for completeness, also w.r.t. the parameters being fixed
in the final analysis~:

\begin{equation}
\label{grad_x_1}
\frac{\partial \Delta q_i}{\partial \eta_i} = 
        \frac{1}{\eta_i} \Delta q_i \;, \quad
\frac{\partial \Delta q_i}{\partial a_i} = 
\left( \log(x) - \frac{1}{T}\frac{\partial T}{\partial a_i} \right)
\Delta q_i \;, \nonumber 
\end{equation}

\begin{equation}
\frac{\partial \Delta q_i}{\partial b_i} = 
\left( \log(1-x) - \frac{1}{T}\frac{\partial T}{\partial b_i} \right)
\Delta q_i \;, \quad
\frac{\partial \Delta q_i}{\partial \gamma_i} = 
\left( \frac{x}{1+\gamma_i x+\rho_i x^{\frac{1}{2}}} - \frac{1}{T} 
\frac{\partial T}{\partial \gamma_i} \right) 
\Delta q_i \;, \nonumber
\end{equation}

\begin{equation}
\frac{\partial \Delta q_i}{\partial \rho_i} = 
\left( \frac{x^{\frac{1}{2}}}{1+\gamma_i x+\rho_i x^{\frac{1}{2}}} - 
\frac{1}{T} \frac{\partial T}{\partial \rho_i} \right) 
\Delta q_i \;, \nonumber
\end{equation}

\noindent
with
\begin{eqnarray}
\label{grad_x_T}
T & = &  B(a_i,b_i+1) \left( 1 + \frac{\gamma_i a_i}{1+a_i+b_i} \right)
+ \gamma_i  B\left(a_i+\frac{1}{2},b_i+1\right)\;,  \\
\frac{\partial T}{\partial a_i} & = & \left[\psi(a_i) - \psi(a_i+b_i+1)
\right] 
         B(a_i,b_i+1) \left( 1 + \frac{\gamma_i a_i}{1+a_i+b_i} \right) 
         +
         B(a_i,b_i+1)
    \\
  &   & \times \left( \frac{\gamma_i a_i}{(1+a_i+b_i)^2} \right) 
        + \left[\psi(a_i+\frac{1}{2})-\psi(a_i+b_i+\frac{3}{2})\right] 
          \rho_i  B\left(a_i+\frac{1}{2},b_i+1\right)\;, \\
\frac{\partial T}{\partial b_i} & = & 
\left[\psi(b_i+1)-\psi(a_i+b_i+1)\right]
         B(a_i,b_i+1) \left( 1 + \frac{\gamma_i a_i}{1+a_i+b_i} \right) -
         B(a_i,b_i+1)
\nonumber \\
  &   & \times \left( \frac{\gamma_i a_i}{(1+a_i+b_i)^2} \right)
        + \left[\psi(b_i+1)-\psi(a_i+b_i+\frac{3}{2}) \right]
          \rho_i  B\left(a_i+\frac{1}{2},b_i+1\right)\;,  \\
\frac{\partial T}{\partial \gamma_i} & = &  B(a_i,b_i+1)
\left( \frac{a_i}{1+a_i+b_i} \right)\;,           \\
\frac{\partial T}{\partial \rho_i}   & = &  
B\left(a_i+\frac{1}{2},b_i+1\right)\;.
\end{eqnarray}

\noindent
Both approaches have been used at the input scale and delivered the
same error contours.
\section{Results of the QCD Analysis}
\label{sec:qcda}
\subsection{Standard QCD Analysis}
\label{sec:qcdastd}

\vspace{2mm}
\noindent
In the fitting procedure we started with the 13 parameters selected, i.e.
three parameters for each of the four polarized parton distribution and 
$\Lambda_{\rm QCD}$ to be determined. For this set of parameters the
sea--quark distribution was assumed to be described according to 
$SU(3)$ flavor symmetry. This assumption is justified for an 
{\it inclusive} data analysis, which we aim on in     this paper, given
the present accuracy of the data, which may be refined in the forthcoming
by
including also semi--inclusive data, which  become more and more precise.
Analyzing the constraints of the
different parameters at $\chi^2_{\rm min}$, it turns out that the
deep--inelastic scattering data do not constrain the four parameters 
$\gamma_{u_v}$, $\gamma_{d_v}$, $b_{\bar{q}}$, and $b_G$ sufficiently
well, since rather large errors for these quantities are obtained.
In the further procedure we fixed these parameters at their values
obtained in the first minimization and chose the ratio
$b_{\overline{q}}/b_G$ as in the unpolarized case~\cite{GRVr}.
Let us note that the latter choice is one possible option. The lack of
constraining power of the present data on the polarized parton densities
has to be stressed, however. Only more precise data in this range
can improve the situation in the future. In the fit we found some tendency
towards a harder  distribution for the  gluon,
although being not significant given
the errors obtained for the large--$x$ parameters.
It also turns out that the small--$x$ parameter for the gluon and the
sea--quark distribution take essentially values which are shifted
relative to each  other as $\alpha_G = \alpha_S + c$ with $c \sim 0.5
\ldots 1$, with some preference according to the value of 
$\chi^2_{\rm min}$. This lead to values of 
$c = 0.9, 0.6$ in LO ({\tt ISET=1,2}) and
$c = 0.9, 0.5$ for the NLO parameterizations ({\tt ISET = 3,4}),
cf. Appendix~1.

The final minimization was carried out under the above conditions and 
determined the remaining eight parameters with their $1\sigma$ errors and
the corresponding covariance matrix elements. Only fits ending with
a positive--definite covariance matrix were accepted. The values and 
errors of these parameters along with those parameters fixed in the
parameterization, cf.~section~5, are summarized for both values of $c$
in LO and NLO in table~2 for the parameters of the non--perturbative
input distributions. The results on $\Lambda_{\rm QCD}$ are discussed
separately in section~6.2. The starting scale of the evolution was
chosen as $Q^2_0 = 4 \GeV^2$. The covariance matrix elements for the
LO and NLO fits, from which  the parameterizations {\tt ISET=1...4}
are derived, are given in tables~3--6.

In Figures~1--4 the fitted parton distribution functions in leading and
next-to-leading order for all sets of            parameterizations and
their 1$\sigma$ errors are presented at the starting scale 
$Q_0^2$.            The positivity bounds as e.g. obtained referring to
the unpolarized distributions~\cite{GRVr} hold either for the central
value itself or well within the present error bands.
The current data constrain the up-valence distribution at best,
followed by the down-valence, sea-quark, and gluon distributions.
Our leading order results deviate from previous parameterizations
\cite{AAC,GRSV}
somewhat at lower values of $x$. Except for the gluon distribution this
deviation is much
 less in next-to-leading order, taking into account the
experimental 1$\sigma$ errors.  In the  range $x~\epsilon~[10^{-3},1]$
the polarized up--valence and gluon distributions are positive within
errors,
while the sea--quark and
the down-valence densities are negative at the reference scale $Q_0^2
= 4 \GeV^2$, the latter
except in a small range at very large $x$.

In Figure~5 the NLO polarized gluon densities of {\tt ISET =3,4}, are
compared to the unpolarized gluon distribution of Ref.~\cite{GRVr},
to which the error of the gluon density as determined
by the H1 experiment~\cite{H1} is overlaid  symmetrically in the range
   $x > 0.1$. Both error contours illustrate the current
margin of the positivity constraint for the gluon density,
                  which is well covered.

The polarized structure function $xg_1^p(x.Q^2)$ measured in the interval
$3.0~\GeV^2 < Q^2 < 5.0~\GeV^2$, Figure~6, using the world asymmetry data
is well described by our QCD NLO curve and the $1\sigma$ error band. We
also compare to corresponding representations of the parameterizations
\cite{AAC,GRSV}, which are compatible within the present errors.

Looking at the $Q^2$ dependence of the structure function
$g_1(x,Q^2)$ in intervals of $x$ gives insight to the scaling violations 
in           the spin sector. As in the unpolarized case the presence of
scaling violations
are expected to manifest in a slope changing with $x$. The world
proton data on $g_1(x,Q^2)$ have been plotted in such a way in
Figure~7 and confronted with the QCD NLO curves of the present
analysis  and its $1\sigma$ error bands. Corresponding curves of the
parameterizations \cite{AAC,GRSV} are also shown.
Slight but non-significant differences between the different analyzes are
observed in the intervals at           low values of $x$.
However, the data are well covered within the
errors by all three analyzes. The current statistics in the low--$x$
range is still rather low.

In Figures~8--11 the scaling violations of the individual polarized
momentum densities are depicted in the range $x~\epsilon~[10^{-3},1],~~
Q^2~\epsilon~[1,10^4]~\GeV^2$ choosing the NLO distributions of 
{\tt ISET=3} as an example. The up-valence distribution $x\Delta u_v$,
Figure~8,
evolves towards smaller values of $x$ and the peak around $x \sim 0.25$
becomes more flat in the evolution from $Q^2=1 \GeV^2$ to $Q^2 = 10^4
\GeV^2$. The distribution is positive within the errors. Statistically 
this distribution is constrained best among all others. The down-valence
distribution $x \Delta d_v$, Figure 9,
remains negative in the same range, although it is
less constraint by the present data than the up-valence density. 
Also here the
evolution is towards smaller values of $x$ and structures at larger
$x$ flatten out. The momentum density of the polarized gluon
$x \Delta G$, Figure~10,
 is positive in the depicted range for $Q^2 \geq 4~\GeV^2$,
but becomes slightly
negative for smaller values of $x$ at $Q^2 \sim 1~\GeV^2$.
Also in  this  case        the evolution moves the shape towards
lower values of $x$ and flattens the distribution. The error band
becomes more uniformly. The sea--quark distribution 
$x \Delta \overline{q}$, Figure~11,
 is negative in the kinematic range shown
for $Q^2 \leq 10 \GeV^2$ and remains negative within errors for $x \leq
5 \cdot 10^-2$ up to $Q^2 = 10^4 \GeV^2$, but changes sign for
larger values of $x$. The minimum of the distribution at $Q^2 = 1 \GeV^2$
around $x \sim 0.1$ moves to $x \sim 0.01$ at $Q^2 = 10^4 \GeV^2$. At
the same time a maximum at $x \sim 0.1$ occurs.

In the present analysis the structure functions $g_1^{p,n}(x,Q^2)$
were parameterized in the twist--2 approximation at next-to-leading order.
As the data may contain higher twist terms         as well it is of
interest to look for the potential effect of these contributions.
At present the size of twist--4 contributions in deep--inelastic
scattering data is widely unknown both in the unpolarized and polarized
case\footnote{For a recent analysis of gluon induced contributions
in the unpolarized case see Ref.~\cite{BRRZ}.}. Due to this we use
the following phenomenological ans\"atze for a higher twist term~:
\begin{eqnarray}
\label{eqHT1}
h_{\rm I}(x,Q^2) &=& 1 + \frac{A}{Q^2} x^\alpha (1-x)^\beta~, \\
\label{eqHT2}
h_{\rm II}(x,Q^2) &=& 1 + \frac{1}{Q^2} \left[A + B x + C x^2\right]~,
\end{eqnarray}
which are used multiplicatively with $g_1(x,Q^2)$. We compare these
fit results with the NLO parameterization {\tt ISET = 4} in Figure~12.
Both ans\"atze yield results of similar size and deviate from the
NLO curves at small values of $Q^2$, however, they are fully consistent
with the NLO results within the
1$\sigma$ error (hatched area). This shows that the present data
do not contain significant higher twist contributions in the range
$Q^2 > 1 \GeV^2$ and a NLO analysis can be carried out. This is also
reflected by the observed logarithmic scaling violations, Figures~7, 12,
and the measurement of the QCD scale $\Lambda_{\rm QCD}$, to which we
turn now.
\subsection{{\boldmath $ \Lambda_{QCD}$} and {\boldmath %
$\alpha_s(M_z)$}}
\label{sec:lamals}

\vspace{2mm}
\noindent
In the QCD analysis we parameterized the strong coupling constant 
$\alpha_s$ in terms of four massless flavors determining 
$\Lambda_{\rm QCD}^{(4)\rm \overline{MS}}$. The NLO result fitting the 
asymmetry data, $g_1/F_1$ or $A_1$, is
\begin{eqnarray}
\Lambda_{\rm QCD}^{(4)\rm \overline{MS}} &=& 235 \pm 53 \;
({\rm stat})\; \MeV,~~{\tt ISET =3}, \\
\Lambda_{\rm QCD}^{(4)\rm \overline{MS}} &=& 240 \pm 60 \;
({\rm stat})\; \MeV,~~{\tt ISET =4},
\end{eqnarray}
identifying both the factorization and renormalization scales with $Q^2$. 
The stability of the NLO result was investigated by changing both scales 
to values different of $Q^2$. Since the present range in $Q^2$ probed in 
polarized deep inelastic scattering is still rather low we vary $Q^2$ only
by factors of 2 and keep the other scale at $Q^2$. The following variations 
are obtained:

\begin{eqnarray}
\Delta \Lambda_{\rm QCD}^{(4)\rm \overline{MS}} =
\begin{array}{c} +  61 \\ -\;47 \end{array} \;~{\rm (fac)}
\begin{array}{c} -  61 \\ + \;114 \end{array} \;~{\rm (ren)} \; \MeV,
~~{\tt ISET = 3}~,
\nonumber   \\
\Delta \Lambda_{\rm QCD}^{(4)\rm \overline{MS}} =
\begin{array}{c} +  58 \\ -\;45 \end{array} \;~{\rm (fac)}
\begin{array}{c} -  66 \\ + \;123 \end{array} \;~{\rm (ren)} \; \MeV,
~~{\tt ISET = 4}~.
\nonumber
\end{eqnarray}

\noindent
These results can be expressed in terms of $\alpha_s(M_Z^2)$:
\begin{eqnarray}
\alpha_s(M_Z^2) = 0.113 \begin{array}{c} + 0.004 \\ - 0.004
\end{array}  
{\rm (stat)}
\begin{array}{c} + 0.004 \\ - 0.004 \end{array}{\rm (fac)}
\begin{array}{c} + 0.008 \\ - 0.005 \end{array}{\rm (ren)}~,
~~{\tt ISET =3},
\nonumber \\
\alpha_s(M_Z^2) = 0.114 \begin{array}{c} + 0.004 \\ - 0.005
\end{array}  
{\rm (stat)}
\begin{array}{c} + 0.004 \\ - 0.004 \end{array}{\rm (fac)}
\begin{array}{c} + 0.008 \\ 
- 0.006 \end{array}{\rm (ren)},~~{\tt ISET =4}~.
\nonumber
\end{eqnarray}
Combining the errors these values
\begin{eqnarray}
\alpha_s(M_Z^2) = 0.113 \begin{array}{c} + 0.010 \\ - 0.008~,
\end{array}  \\
\alpha_s(M_Z^2) = 0.114 \begin{array}{c} + 0.010 \\ - 0.009~,
\end{array}  
\end{eqnarray}
can be compared with results from other QCD analyzes of polarized 
inclusive deep--inelastic scattering data

\vspace*{-0.25cm}
\begin{eqnarray}
{\rm E154}~\cite{E154QCD}: \quad \alpha_s(M_Z^2) & = & 0.108 -
0.116\,, \quad {\rm bad \,~for} \ge 0.120~,  \nonumber \\
{\rm SMC}~\cite{SMCQCD}: \quad \alpha_s(M_Z^2) & = & 0.121 \pm 0.002
{\rm (stat)} \pm 0.006 {\rm (syst+theor)}~, \nonumber \\
{\rm ABFR}~\cite{ABFR}: \quad \alpha_s(M_Z^2) & = & 0.120
\begin{array}{c} + 0.004 \\ - 0.005 
\end{array} {\rm (exp)}
\begin{array}{c} + 0.009 \\ - 0.006 \end{array} {\rm (theor)}~,
\end{eqnarray}

\noindent
and with the value of the current world average 
\begin{eqnarray}
 \alpha_s(M_Z^2) = 0.118 \pm 0.002~~\cite{PDG}~.
\end{eqnarray}
The results of the present analysis are consistent within 1$\sigma$
although our result and that of E154~\cite{E154QCD} lead to a 
somewhat lower central value for $\alpha_s(M_Z^2)$. In the scheme--invariant 
analysis, which allows for an alternative view, a minor shift for 
$\Lambda_{\rm QCD}^{(4)\rm \overline{MS}}$ down by $12\;\MeV$ was 
found leading to a central value of $\alpha_s(M_Z^2) = 0.113$ with 
statistical and the renormalization scale dependent errors of the same 
size as quoted above. Still more systematic effects have to be 
investigated in the future. Here we see a main point in studying further 
the unpolarized structure function $F_1(x,Q^2)$, the denominator function
in the expression for the spin asymmetry. In our analysis it was obtained
from parameterizations of $F_2$~\cite{F2NMC} and $R$~\cite{R1990} 
measurements which themselves are subject to systematic uncertainties.
\subsection{Scheme-Invariant Analysis}
\label{sec:qcdsi}

\vspace{2mm}
\noindent
A factorization--scheme invariant QCD analysis in next--to--leading
order based on the observables $g_1(x,Q^2)$ and $\partial g_1(x,Q^2) /
\partial t$ for the  proton has been performed, where $t$ is the
evolution variable defined in Eq.~(\ref{eqT}), see section~\ref{sec:si}. 
If compared to the standard analysis these two observables take the roles
of $\Delta \Sigma(x,Q^2)$ and $\Delta G(x,Q^2)$.
Such an analysis has, in principle,  the advantage of direct experimental
control over the input densities since they are {\it measurable} 
quantities.
In this way no ansatz for
$\Delta G$ is necessary and the only free parameter to be determined
is $\Lambda_{\rm QCD}$. Unfortunately, the quality of the present data
does not yet allow an experimental
determination of the slope $\partial g_1(x,Q^2) /
\partial t$ accurate enough to be used as input density. In the
present analysis this  slope
has been derived  fitting the world data as described in section~3.
In the singlet analysis the initial distributions are $g_1^S(x,Q^2_0)$
and                                                   
$\partial g_1^S(x,Q^2_0)/\partial t$. The latter quantity is depicted
in Figure~13 as a function of $x$ together with the corresponding
slopes at higher values of $Q^2$.
This slope has a rather involved shape and it requires precise
data in the range of the input scale $Q_0^2$ on $g_1(x,Q^2)$ 
to determine the slope experimentally in the future.
The QCD analysis performed lead to
a downward shift of $12~\MeV$ for
$\Lambda_{\rm QCD}$ which yields a similar result
for $\alpha_s(M_Z^2)$ as obtained in the standard analysis, 
see section~ 6.2.
\section{Moments of Polarized Parton Distributions}
\label{sec:moments}

\vspace{2mm}
\noindent
In recent lattice simulations~[23--25]
low moments for the
polarized parton densities $\Delta u_v, \Delta d_v$, and $\Delta u -
\Delta d$ were determined. These moments are 
\begin{eqnarray}
\langle x^{n+1} \rangle_{\Delta f} = \int_0^1 x^{n+1} \left[\Delta f(x)
+ (-1)^{(n+1)} \Delta
\overline{f}(x)\right] dx  = \Mvec[\Delta f(x)](n) 
+ (-1)^{(n+1)}
\Mvec[\Delta \overline{f}(x)](n)~,
\end{eqnarray}
for $n=-1,0,1$.
The results of the simulation are presented at a scale $\mu^2 =1/a^2
\sim 4~\GeV^2$, where $a$ denotes the lattice spacing. The values of
these lattice moments may be compared with the moments being obtained
in the present analysis. Moreover we also present the respective
moments for the sea--quark and gluon distributions and
include their $1\sigma$  error.
Moments of structure functions or parton distributions
based on a data analysis are only constrained in the range of the
respective measurements. No definite statement can be made on the
behavior of the (non-perturbative) parton distributions outside the
range being explored.\footnote{Note that before 1992 the strong growth
of the structure function $F_2(x,Q^2)$, later found even at very low 
scales $Q^2 \sim 2 \GeV^2$, came as a surprise~\cite{DIS92}. Likewise,
at  present  the unpolarized gluon and sea--quark
distributions at large values of $x$ are widely unknown.} Reliably
precise estimates neither for the range of very low~\cite{SX} nor
large $x$ are available currently. Due to this the value and error
of the moments given below is that derived from the current experimental
constraints. Nonetheless we study which contribution of the moments
is determined by the range in which measurements exist.

In Table~7 the moments of the NLO parton densities and some of their
combinations, being derived in the present analysis, and their $1\sigma$
errors are given. We also compare to the respective values obtained
integrating only over the kinematic range in which data are measured
and quote the value below and above the kinematic range for the
moments of the NLO parton densities {\tt ISET=3,4}.
The determination of the error of the moments requires in general
a correlated error propagation through the evolution equations, if the
reference scale $Q^2$ is not the input scale $Q_0^2$. As the lattice 
calculations provide measurements for a series of moments at the starting
scale chosen in the present analysis the formulae given in section~5.2.
apply directly. Whereas the estimated uncertainty for the distributions
{\tt ISET = 3,4} for the unmeasured large-$x$ range $x > 0.85$ is small,
it is quite significant for the lowest moment for the small--$x$ range,
$x < 0.02$. For the up--quark distributions and their
combinations the present correlated 1$\sigma$  errors are about 10\%
for the lowest moments, and reach about 30\% for the down quark, and
25\% for the sea quarks. Depending on the parameterization the
lowest moment of the polarized gluon distribution is found to have an
error of 50\% and more. Towards higher moments the errors grow, but the
small--$x$ uncertainties become less significant. The first three
moments of the distributions $\Delta u_v(x), \Delta d_v(x)$ and
$\Delta u(x) - \Delta d(x)$ can be compared with lattice results~[23--25],
with some caution. Only for a part of these values continuum 
extrapolations were yet performed. Still further study is required
to safely determine and quantify systematic errors.
While for the distribution $\Delta u - \Delta d$ the moment for $n=-1$
comes out to be too small in the lattice measurements, it is somewhat
too large for the moments $n=0,1$. We quote for the lowest
moment an error, which is obtained due to all correlated parameters
in the fit. On the other hand, the central value given is the axial
charge $g_A$, which is much more precisely determined by other 
measurements. It is yet difficult  to see a systematic trend in
the lattice values comparing to the moments found in the present 
analysis. Given the fact, that higher moments are
more difficult to measure the agreement of the first two moments within
the errors determined in the present analysis is quite good.

We also compare the first moments in NLO with other recent analyzes, see 
table~8. Although some of the numbers, particularly for the gluon and
sea--quark distributions look different they agree perfectly within
the  1$\sigma$ errors derived in our analysis.
\section{Conclusions}
\label{sec:conc}

\vspace{2mm}
\noindent
We have performed a QCD analysis of the inclusive polarized 
deep--inelastic charged lepton--nucleon scattering world
data to next--to--leading order and derived    parameterizations of
polarized parton distributions at a starting scale $Q_0^2$ together with
the QCD--scale $\Lambda_{\rm QCD}$. The  analysis was performed using the
$\chi^2$--method to determine the parameters of the problem in a fit
to the data. 
A new aspect in comparison with previous analyzes is that
we determine also the fully correlated errors of the parton densities and
the QCD scale in leading and next--to--leading order.
Due to the fact that not all shape parameters of the
parton densities can be measured at sufficient accuracy using the
present data, we derived two sets of parameterizations, which mainly 
differ in the low--$x$  behavior of the gluon densities.  Detailed
comparisons were performed to the results obtained in other recent
prameterizations~\cite{AAC, GRSV}. The previous results are widely
compatible with the present parameterizations within the current 
1$\sigma$ error bands.
Since we used the {\sc Mellin}--method to solve the evolution equations
the Gaussian error propagation of the parameters of the input densities
through the evolution was possible in analytic form.
Both the central values and the 1$\sigma$ errors of the
parton densities are made available in form of a numerical
parameterizations in the kinematic range $x~\epsilon~[10^{-9},1],~~
Q^2~\epsilon~[1,10^6]~\GeV^2$.  These distributions can be used 
in the numerical calculations for polarized high--energy scattering
processes at hadron-- and $ep$--colliders. Moreover, due to the fact
that parameterizations of the errors are available, error estimates
of these quantities are possible w.r.t. the present knowledge of
parton densities. These parameterizations are available as fast
{\tt FORTRAN}--routines which makes their application possible
in Monte Carlo simulations.

The current experimental data allow to measure the QCD scale 
$\Lambda_{\rm QCD}$ with a statistical error of $\delta \Lambda = \pm
60 \MeV$, and $\delta \alpha_s(M_Z^2) \pm 0.004$. 
Since at present only a next-to-leading order analysis
can be carried out the variations in the renormalization and 
factorization scales induce yet large systematic errors. Combining
all errors one obtains for the two scenarios ({\tt ISET = 3,4})
$\alpha_s(M_Z^2) = 0.113 \pm {\footnotesize \begin{array}{c} 0.010\\
0.008 \end{array}}$ and
$\alpha_s(M_Z^2) = 0.114 \pm {\footnotesize \begin{array}{c} 0.010\\
0.009 \end{array}}$, respectively. The theoretical error due to the
present renormalization and factorization scale uncertainties can be
further reduced in a three--loop analysis, for which the anomalous
dimensions have still to be calculated.
We also
performed for the first time a factorization scheme--invariant
QCD analysis based on $g_1(x,Q^2)$ and $\partial g_1(x,Q^2)/\partial
\log Q^2$ which lead to a small shift of $12 \MeV$ in $\Lambda_{\rm
QCD}$ only. This novel way of analysis which is based on the scaling
violations of {\it observables} directly may show its full strengths
in later analyzes based on the even higher statistics of future
experiments. In this analysis factorization scale uncertainties do not
occur.

The results of the present analysis may be compared to recent lattice
results calculating the first few moments of the distributions
$\Delta u_v(x,Q^2)$, $\Delta d_v(x,Q^2)$, and $\Delta (u_v-d_v)(x,Q^2)$ 
and their respective errors. We also present the respective moments
of $\Delta \overline{q}(x,Q^2)$ and $\Delta G(x,Q^2)$ for which lattice
results do not exist at present. Both on the side of the lattice 
measurements and the extraction of the distribution functions from deep 
inelastic scattering data the errors improved during recent years and the
values became closer. However, more work has yet to be done in the future
on systematic effects and even more precise experimental data would be 
welcome to perform an essential test of QCD also in this field at a 
higher precision than possible at present.


\vspace{2mm}
\noindent
{\bf Acknowledgment}.~This work was supported in part by EU contract
FMRX-CT98-0194 (DG 12 - MIHT). For discussions in an early phase of this
work we would like to thank Andreas Vogt. For discussions we thank
W.-D. Nowak. We thank Stefano Forte and Giovanni Ridolfi for recalculating 
the first moments from their previous analyzes in the $\overline{\rm MS}$ 
scheme. For discussion on recent lattice results we thank S.~Capitani, 
K.~Jansen, and G.~Schierholz.

\vspace{2mm}
\noindent

\newpage
\section{Appendix: The {\tt FORTRAN}-code for the parton densities and
their errors}

\vspace{2mm}
\noindent
A fast {\tt FORTRAN} program is available to represent the polarized
parton densities $x\Delta u_v(x,Q^2)$, $x\Delta d_v(x,Q^2)$, $x\Delta
G(x,Q^2)$, and $x\Delta \bar{q}(x,Q^2)$ and the polarized structure
functions $xg_1^p(x,Q^2)$ and $xg_1^n(x,Q^2)$ in leading and
next-to-leading order in the $\overline{\rm MS}$--scheme together with
the parameterizations of their $1\sigma$  errors.      The
following ranges in $x$ and $Q^2$ are covered:   

\vspace*{-0.15cm}

\begin{center}
$10^{-9} < x < 1 \quad , \quad 1~\GeV^2 < Q^2 < 10^6~\GeV^2.$
\end{center} 

\vspace*{-0.15cm}

\noindent
The polarized distributions are the result of a fit to the world data
on spin asymmetries, i.e. $A_1^{p,n,d}$ or $g_1/F_1^{p,n,d}$, as
described in the paper. 
The {\tt SUBROUTINE PPDF} returns the values of the polarized
distributions, always multiplied with $x$, at a given point in
$x$ and $Q^2$ by interpolating the data on specified grids. The
interpolation in $x$ is done by cubic splines and in $Q^2$ by a
linear interpolation in $\log\,(Q^2)$.~\footnote{We thank S.~Kumano and
M.~Miyama of the AAC--collaboration for allowing us to use their 
interpolation routines.}

The parton distributions are evaluated by
\begin{center}
{\tt 
SUBROUTINE PPDF(ISET, X, Q2, UV, DUV, DV, DDV, GL, DGL, SEA, DSEA, G1P,%
DG1P,G1N,DG1N)}
\end{center}
All non-integer variables are of the type {\tt REAL*8}. The calling
routine has to contain the {\tt COMMON/INTINI/ IINI}. Before the first 
call to {\tt SUBROUTINE PPDF} the initialization is set by {\tt IINI = 0}.
The values of the parameter {\tt ISET} are:
\begin{eqnarray}
{\tt ISET = 1} & & {\rm ~~LO},~~~\alpha_G = \alpha_S+0.9 \nonumber\\
{\tt ISET = 2} & & {\rm ~~LO},~~~\alpha_G = \alpha_S+0.6 \nonumber\\
{\tt ISET = 3} & & {\rm NLO},~~~\alpha_G = \alpha_S+1.0 \nonumber\\
{\tt ISET = 4} & & {\rm NLO},~~~\alpha_G = \alpha_S+0.5 \nonumber
\end{eqnarray}
The parameters {\tt X, Q$^2$/GeV$^2$} are $x$ and $Q^2$. The momentum
densities of the polarized up- and down valence quarks, gluons and
the sea quarks are {\tt UV, DV, GL, SEA}, with ${\tt SEA}~=~x\Delta
u_s = x\Delta d_s = x\Delta \overline{u} = x\Delta \overline{d}
= x\Delta s = x\Delta \overline{s}$. Correspondingly, {\tt DUV}
is the $1\sigma$  error of {\tt UV} etc. and {\tt G1P} and {\tt G1N}
are the values of the electromagnetic structure functions $g_1^p$
and $g_1^n$.

\noindent
The program can be received on request via e-mail to 
{\tt Johannes.Bluemlein@desy.de} or {\tt Helmut.Boettcher@desy.de} or
from {\tt http://www-zeuthen.desy.de/$\sim$hboett/ppdf.uu.gz}.
\newpage

\section{Tables}
\label{sec:tab}

\vspace*{2mm} \noindent
%
\renewcommand{\arraystretch}{1.3}
%
\normalsize
\begin{center}
\begin{tabular}{|l|c|c|r|r|c|}
\hline \hline 
Experiment & $x$--range & $Q^2$--range  & \multicolumn{2}{c|}{number
of data points} & Ref. \\ \cline{4-5}
           &            & [$\GeV^2$]     
           & $g_1/F_1$ or $A_1$ & \multicolumn{1}{c|}{$g_1$} & \\
\hline \hline
E143(p)       & 0.027 -- 0.749 & 1.17 -- 9.52  &  82 &  28 & 
\cite{E143pd} \\ 
HERMES(p)     & 0.028 -- 0.660 & 1.13 -- 7.46  &  39 &  39 &
\cite{HERMp} \\ 
E155(p)       & 0.015 -- 0.750 & 1.22 -- 34.72 &  24 &  24 &
\cite{E155p} \\ 
SMC(p)        & 0.005 -- 0.480 & 1.30 -- 58.0  &  59 &  12 &
\cite{SMCpd} \\ 
EMC(p)        & 0.015 -- 0.466 & 3.50 -- 29.5  &  10 &  10 &
\cite{EMCp} \\ 
\hline
proton        &                &               & 214 & 113 & \\
\hline \hline
E143(d)       & 0.027 -- 0.749 & 1.17 -- 9.52  &  82 &  28 &
\cite{E143pd} \\ 
E155(d)       & 0.015 -- 0.750 & 1.22 -- 34.79 &  24 &  24 &
\cite{E155d} \\ 
SMC(d)        & 0.005 -- 0.479 & 1.30 -- 54.8  &  65 &  12 &
\cite{SMCpd} \\ 
\hline
deuteron      &                &               & 171 &  64 & \\
\hline \hline
E142(n)       & 0.035 -- 0.466 & 1.10 -- 5.50  &  30 &   8 &
\cite{E142n} \\ 
HERMES(n)     & 0.033 -- 0.464 & 1.22 -- 5.25  &   9 &   9 &
\cite{HERMn} \\ 
E154(n)       & 0.017 -- 0.564 & 1.20 -- 15.0  &  11 &  17 &
\cite{E154n}/\cite{E154QCD} \\ 
\hline
neutron       &                &               &  50 &  34 & \\
\hline \hline
total         &                &               & 435 & 211 & \\  
\hline \hline            
\end{tabular}
\end{center}
\normalsize

\vspace{2mm}
\noindent
\begin{center}
{\sf Table~1: Published data points above $Q^2 = 1.0~\GeV^2$.}
\end{center}
\renewcommand{\arraystretch}{1}
%
\newpage
\renewcommand{\arraystretch}{1.3}
%
\begin{center}
\begin{tabular}{|c|c|c|c|c||c|c|c|c|}
\hline \hline
           & \multicolumn{4}{c||}{Scenario 1} & 
             \multicolumn{4}{c|}{Scenario 2} \\
\hline
           & \multicolumn{2}{c|}{LO} & \multicolumn{2}{c||}{NLO} & 
             \multicolumn{2}{c|}{LO} & \multicolumn{2}{c|}{NLO} \\
           & \multicolumn{2}{c|}{{\tt ISET=1}} & \multicolumn{2}{c||}
           {{\tt ISET=3}} & 
             \multicolumn{2}{c|}{{\tt ISET=2}} & \multicolumn{2}{c|}
             {{\tt ISET=4}} \\
\cline{2-9} 
           & value & error & value & error & value & error & value & error \\ 
\hline \hline
$\Lambda_{QCD}^{(4)}, \MeV$ & 203 & 120 & 235 & 53 & 195 & 143 & 240 & 60 \\
\hline \hline
$\eta_{u_v}$      &  0.926 & fixed &  0.926 & fixed & 0.926 & fixed &
0.926 & fixed \\
$a_{u_v}$         &  0.197 & 0.013 &  0.294 & 0.035 & 0.199 & 0.013 &
0.271 & 0.029 \\
$b_{u_v}$         &  2.403 & 0.107 &  3.167 & 0.212 & 2.416 & 0.107 &
3.070 & 0.175 \\
$\gamma_{u_v} (*)$ & 21.34 & fixed & 27.22  & fixed & 21.34 & fixed &
27.22 & fixed \\
\hline \hline
$\eta_{d_v}$      & -0.341 & fixed & -0.341 & fixed & -0.341 & fixed &
-0.341 & fixed  \\
$a_{d_v}$         &  0.190 & 0.049 &  0.254 & 0.111 & 0.182 & 0.046 &
0.325 & 0.125 \\
$b_{d_v}$         &  3.240 & 0.884 &  3.420 & 1.332 & 3.209 & 0.895 &
3.925 & 1.129 \\
$\gamma_{d_v} (*)$ & 30.80 & fixed & 19.06  & fixed & 30.80 & fixed &
19.06 & fixed \\
\hline \hline
$\eta_{sea}$      & -0.353 & 0.054 & -0.447 & 0.082 & -0.314 & 0.053 &
-0.435 & 0.061\\
$a_{sea}$         &  0.367 & 0.048 &  0.424 & 0.062 & 0.428 & 0.055 &
0.285 & 0.048 \\
$b_{sea} (*)$     &  8.51  & fixed &  8.93  & fixed & 8.51 & fixed &
8.93 & fixed \\
\hline \hline
$\eta_G$          &  1.281 & 0.816 &  1.026 & 0.554 & 1.043 & 0.938 &
0.931 & 0.679 \\
\cline{2-9}
$a_{G}$           & \multicolumn{2}{c|}{$a_{sea} + 0.9$} &  
                    \multicolumn{2}{c||}{$a_{sea} + 1.0$}& 
           \multicolumn{2}{c|}{$a_{sea} + 0.6$} &  
                    \multicolumn{2}{c|}{$a_{sea} + 0.5$} \\
\cline{2-9}
$b_{G} (*)$       &  5.91  & fixed &  5.51  & fixed & 5.91 & fixed &
5.51 & fixed \\
\hline \hline
$\chi^2$ / NDF    & \multicolumn{2}{c|}{1.02}  &
                    \multicolumn{2}{c||}{0.90} & 
                    \multicolumn{2}{c|}{1.04}  &
                    \multicolumn{2}{c||}{0.93} \\
\hline \hline
\end{tabular}
\end{center}
\normalsize

\vspace*{2mm}
\noindent
{\sf Table~2: Parameter values in LO and NLO 
($\overline{\rm MS}$) of the $7+1$
parameter fit based on the world asymmetry data for both
scenarios. The $(*)$ marks those parameters which were fixed 
after the first minimization
since the present data do not constrain these parameters well enough
(see text).   
 }
\renewcommand{\arraystretch}{1}
%
\newpage
%
\renewcommand{\arraystretch}{1.3}
\begin{center}
\begin{tabular}{||c||c|c|c|c|c|c|c|c||}
\hline \hline
\multicolumn{9}{||c||}{LO}\\
\hline \hline
        & $\Lambda_{QCD}^{(4)}$ & $a_{u_v}$ & $b_{u_v}$ & $a_{d_v}$ & $b_{d_v}$ & $\eta_{sea}$ & $a_{sea}$ & $\eta_G$ \\
\hline \hline
 $\Lambda_{QCD}^{(4)}$ &  1.43E-2 &  &  &  &  &  &  &  \\
\hline
 $a_{u_v}$    & -2.05E-5 &  1.80E-4 &  &  &  &  &  &  \\
\hline
 $b_{u_v}$    & -9.07E-5 &  3.91E-4 &  1.15E-2 &  &  &  &  &  \\
\hline
 $a_{d_v}$    &  1.10E-4 &  1.03E-5 & -2.40E-3 &  2.43E-3 &  &  &  &  \\
\hline
 $b_{d_v}$    & -4.65E-5 & -7.92E-3 & -6.86E-3 &  5.48E-3 &  7.82E-01 &  &  &  \\
\hline
 $\eta_{sea}$ &  1.02E-4 & -4.46E-4 & -2.84E-3 &  9.85E-4 &  2.82E-2 &  2.94E-3 &  &  \\
\hline
 $a_{sea}$    & -4.31E-5 &  1.58E-4 &  1.33E-3 & -5.96E-4 & -9.32E-3 & -2.58E-4 &  2.29E-3 &  \\
\hline
 $\eta_G$     & -1.03E-3 &  2.02E-3 &  1.58E-2 & -2.78E-3 & -1.61E-1 & -1.59E-2 &  9.56E-3 &  6.65E-1 \\
\hline
\hline
\end{tabular} 
\end{center}
\normalsize
\vspace*{+2mm}
\noindent
{\sf Table~3: The covariance matrix for the $7+1$ parameter LO fit
(scenario 1, {\tt ISET = 1}) based on the world asymmetry data.} 
\renewcommand{\arraystretch}{1.0}
%

\vspace*{2cm}
%
\renewcommand{\arraystretch}{1.3}
%
\begin{center}
\begin{tabular}{||c||c|c|c|c|c|c|c|c||}
\hline \hline
\multicolumn{9}{||c||}{NLO}\\
\hline \hline 
        & $\Lambda_{QCD}^{(4)}$ & $a_{u_v}$ & $b_{u_v}$ & $a_{d_v}$ & $ b_{d_v}$ & $\eta_{sea}$ & $a_{sea}$ & $\eta_G$ \\
\hline \hline
 $\Lambda_{QCD}^{(4)}$ &  2.81E-3 &  &  &  &  &  &  & \\
\hline
 $a_{u_v}$    &  2.71E-5 &  1.22E-3 &  &  &  &  &  & \\
\hline
 $b_{u_v}$    & -1.30E-4 &  5.10E-3 &  4.50E-2 &  &  &  &  & \\
\hline
 $a_{d_v}$    & -3.35E-4 & -5.17E-4 & -3.23E-3 &  1.23E-2 &  &  &  & \\
\hline
 $b_{d_v}$    & -6.22E-4 & -1.27E-2 &  4.65E-2 &  8.29E-2 &  1.78E-0 &  &  & \\
\hline
 $\eta_{sea}$ & -5.30E-5 & -2.13E-3 & -1.12E-2 &  5.19E-3 &  4.74E-2 &  6.77E-3 &  & \\
\hline
 $a_{sea}$    & -4.85E-6 &  9.07E-4 &  4.49E-3 & -3.78E-3 & -2.98E-2 & -2.39E-3 &  3.82E-3 & \\
\hline
 $\eta_G$     &  4.03E-4 &  1.41E-2 &  6.71E-2 & -3.07E-2 & -2.22E-1 & -3.78E-2 &  1.90E-2 &  3.07E-1 \\
\hline
\hline
\end{tabular} 
\end{center}
\normalsize
\vspace{+2mm}
\noindent
{\sf Table~4: The covariance matrix for the $7+1$ parameter NLO fit
(scenario 1, {\tt ISET = 3}) based on the world asymmetry data.} 
\renewcommand{\arraystretch}{1.0}
%
\newpage
%
\renewcommand{\arraystretch}{1.3}
\begin{center}
\begin{tabular}{||c||c|c|c|c|c|c|c|c||}
\hline \hline
\multicolumn{9}{||c||}{LO}\\
\hline \hline
        & $\Lambda_{QCD}^{(4)}$ & $a_{u_v}$ & $b_{u_v}$ & $a_{d_v}$ & $b_{d_v}$ & $\eta_{sea}$ & $a_{sea}$ & $\eta_G$ \\
\hline \hline
 $\Lambda_{QCD}^{(4)}$ &  2.05E-2 &  &  &  &  &  &  &  \\
\hline
 $a_{u_v}$    &  1.25E-4 &  1.76E-4 &  &  &  &  &  &  \\
\hline
 $b_{u_v}$    & -1.37E-3 &  3.42E-4 &  1.15E-2 &  &  &  &  &  \\
\hline
 $a_{d_v}$    & -5.57E-4 &  1.89E-6 & -2.19E-3 &  2.14E-3 &  &  &  &  \\
\hline
 $b_{d_v}$    &  9.62E-3 & -7.84E-3 & -1.29E-3 &  4.74E-3 &  8.00E-01 &  &  &  \\
\hline
 $\eta_{sea}$ &  4.37E-4 & -4.34E-4 & -2.88E-3 &  1.01E-3 &  2.73E-2 &  2.86E-3 &  &  \\
\hline
 $a_{sea}$    & -5.78E-4 &  6.91E-5 &  4.56E-4 & -3.73E-4 & -2.16E-3 &  3.98E-4 &  3.06E-3 &  \\
\hline
 $\eta_G$     &  9.43E-4 &  1.91E-3 &  1.54E-2 & -1.85E-3 & -1.55E-1 & -1.38E-2 &  2.97E-3 &  8.80E-1 \\
\hline
\hline
\end{tabular} 
\end{center}
\normalsize
\vspace*{+2mm}
\noindent
{\sf Table~5: The covariance matrix for the $7+1$ parameter LO fit
(scenario 2, {\tt ISET = 2}) based on the world asymmetry data.} 
\renewcommand{\arraystretch}{1.0}
%

\vspace*{2cm}
%
\renewcommand{\arraystretch}{1.3}
%
\begin{center}
\begin{tabular}{||c||c|c|c|c|c|c|c|c||}
\hline \hline
\multicolumn{9}{||c||}{NLO}\\
\hline \hline 
        & $\Lambda_{QCD}^{(4)}$ & $a_{u_v}$ & $b_{u_v}$ & $a_{d_v}$ & $ b_{d_v}$ & $\eta_{sea}$ & $a_{sea}$ & $\eta_G$ \\
\hline \hline
 $\Lambda_{QCD}^{(4)}$ &  3.55E-3 &  &  &  &  &  &  & \\
\hline
 $a_{u_v}$    &  3.09E-4 &  8.50E-4 &  &  &  &  &  & \\
\hline
 $b_{u_v}$    &  1.08E-3 &  3.88E-3 &  3.06E-2 &  &  &  &  & \\
\hline
 $a_{d_v}$    &  1.88E-3 &  2.66E-4 & -8.80E-4 &  1.56E-2 &  &  &  & \\
\hline
 $b_{d_v}$    &  1.74E-2 & -7.57E-3 & -1.34E-2 &  9.03E-2 &  1.27E-0 &  &  & \\
\hline
 $\eta_{sea}$ &  5.77E-4 & -9.03E-4 & -5.32E-3 &  3.26E-3 &  3.84E-2 &  3.74E-3 &  & \\
\hline
 $a_{sea}$    & -5.33E-4 &  6.81E-4 &  3.68E-3 & -2.28E-3 & -2.56E-2 & -9.06E-4 &  2.34E-3 & \\
\hline
 $\eta_G$     & -8.19E-3 &  1.17E-2 &  5.61E-2 & -3.95E-2 & -3.74E-1 & -2.37E-2 &  1.51E-2 &  4.61E-1 \\
\hline
\hline
\end{tabular} 
\end{center}
\normalsize
\vspace{+2mm}
\noindent
{\sf Table~6: The covariance matrix for the $7+1$ parameter NLO fit
(scenario 2, {\tt ISET=4}) based on the world asymmetry data.} 
\renewcommand{\arraystretch}{1.0}
%
\newpage
\small
%
\renewcommand{\arraystretch}{1.3}
\begin{center}
\begin{tabular}{|l|r|r|c|r|c|r|r|}
\hline \hline 
\multicolumn{1}{|l|}{ }&
\multicolumn{1}{c|}{ }&
\multicolumn{2}{c|}{\tt ISET=3 }&
\multicolumn{2}{c|}{\tt ISET=4 } &
\multicolumn{2}{c|}{ lattice results   } \\
\cline{3-8}
\multicolumn{1}{|c|}{
}& 
\multicolumn{1}{ c|}{$n$} &
\multicolumn{1}{ c|}{value} &
\multicolumn{1}{ c|}{value out} &
\multicolumn{1}{ c|}{value} &
\multicolumn{1}{ c|}{value out} &
\multicolumn{1}{ c|}{QCDSF    } &
\multicolumn{1}{ c|}{LHPC/    } \\
\multicolumn{1}{|c|}{            }&
\multicolumn{1}{ c|}{   } &
\multicolumn{1}{ c|}{     } &
\multicolumn{1}{ c|}{of range } &
\multicolumn{1}{ c|}{     } &
\multicolumn{1}{ c|}{of range } &
\multicolumn{1}{ c|}{         } &
\multicolumn{1}{ c|}{SESAM    } \\
\hline\hline
$\Delta u_v$  &--1 & $0.926 \pm 0.071$
                   & {\footnotesize $0.191|4\EM4$}
                   & $0.926 \pm 0.062$
                   & {\footnotesize $0.207|4\EM4$}
                   & 0.889(29)
                   & 0.860(69)
\\
              &  0 & $0.163 \pm 0.014$
                   & {\footnotesize $0.001|3\EM4$}
                   & $0.160 \pm 0.011$
                   & {\footnotesize $0.001|4\EM4$}
                   & 0.198(8)
                   & 0.242(22)
\\
              &  1 & $0.055 \pm 0.006$
                   & {\footnotesize $1\EM5|3\EM4$}
                   & $0.055 \pm 0.005$
                   & {\footnotesize $1\EM5|3\EM4$}
                   & 0.041(9)
                   & 0.116(42)
\\
              &  2 & $0.024 \pm 0.003$
                   & {\footnotesize $0|3\EM4$}
                   & $0.024 \pm 0.003$
                   & {\footnotesize $0|3\EM4$}
                   &
                   &
\\
\hline
$\Delta d_v$  &--1 & $-0.341 \pm 0.123$
                   & {\footnotesize $-0.104|$-$7\EM5$}
                   & $-0.341 \pm 0.103$
                   & {\footnotesize $-0.086|$-$3\EM5$}
                   & -0.236(27)
                   & -0.171(43)
\\
              &  0 & $-0.047 \pm 0.021$
                   & {\footnotesize $-5\EM4|$-$6\EM5$}
                   & $-0.049 \pm 0.017$
                   & {\footnotesize $-5\EM4|$-$3\EM5$}
                   & -0.048(3)
                   & -0.029(13)
\\
              &  1 & $-0.015 \pm 0.009$
                   & {\footnotesize $-1\EM5|$-$5\EM5$}
                   & $-0.015 \pm 0.007$
                   & {\footnotesize $-1\EM5|$-$2\EM5$}
                   & -0.028(2)
                   &  0.001(25)
\\
              &  2 & $-0.006 \pm 0.005$
                   & {\footnotesize $0|$-$5\EM5$}
                   & $-0.006 \pm 0.004$
                   & {\footnotesize $0|$-$2\EM5$}
                   &
                   &
\\
\hline
$\Delta u$--$\Delta d$
              &--1 & $1.267  \pm 0.142$
                   & {\footnotesize  $0.295|5\EM4$}
                   & $1.267  \pm 0.121$
                   & {\footnotesize  $0.293|5\EM4$}
                   &  1.14(3)
                   &  1.031(81)
\\
              &  0 & $0.210  \pm 0.025$
                   & {\footnotesize  $0.001|4\EM4$}
                   & $0.209 \pm 0.021$
                   & {\footnotesize  $0.001|4\EM4$}
                   &  0.246(9)
                   &  0.271(25)
\\
              &  1 & $0.070 \pm 0.011 $
                   & {\footnotesize $2\EM5|4\EM4$}
                   & $0.069  \pm 0.009$
                   & {\footnotesize $2\EM5|4\EM4$}
                   &  0.069(9)
                   &  0.115(49)
\\
              &  2 & $0.030 \pm 0.006 $
                   & {\footnotesize $0|3\EM4$}
                   & $0.030  \pm 0.004$
                   & {\footnotesize $0|3\EM4$}
                   &
                   &
\\
\hline
$\Delta u$
              &--1 & $0.851  \pm 0.075$
                   & {\footnotesize  $0.152|4\EM4$}
                   & $0.854  \pm 0.066$
                   & {\footnotesize  $0.158|4\EM4$}
                   &
                   &
\\
              &  0 & $0.160  \pm 0.014$
                   & {\footnotesize  $8\EM4|3\EM4$}
                   & $0.158 \pm 0.012$
                   & {\footnotesize  $8\EM4|4\EM4$}
                   &
                   &
\\
              &  1 & $0.055 \pm 0.006 $
                   & {\footnotesize $1\EM5|3\EM4$}
                   & $0.055  \pm 0.005$
                   & {\footnotesize $1\EM5|3\EM4$}
                   &
                   &
\\
              &  2 & $0.024 \pm 0.003 $
                   & {\footnotesize $0|3\EM4$}
                   & $0.024  \pm 0.003$
                   & {\footnotesize $0|3\EM4$}
                   &
                   &
\\
\hline
$\Delta d$
              &--1 & $-0.415 \pm 0.124$
                   & {\footnotesize  -$0.144|$-$7\EM5$}  
                   & $-0.413  \pm 0.104$
                   & {\footnotesize -$0.135|$-$3\EM5$}
                   &
                   &
\\
              &  0 & $-0.050  \pm 0.022$
                   & {\footnotesize -$7\EM4|$-$6\EM5$}
                   & $-0.051 \pm 0.017$
                   & {\footnotesize -$7\EM4|$-$3\EM5$}
                   &
                   &
\\
              &  1 & $-0.015 \pm 0.009 $
                   & {\footnotesize -$1\EM5|$-$5\EM5$}
                   & $-0.015  \pm 0.007$
                   & {\footnotesize -$1\EM5|$-$2\EM5$}
                   &
                   &
\\
              &  2 & $-0.006 \pm 0.005 $
                   & {\footnotesize $0|$-$5\EM5$}
                   & $-0.006  \pm 0.004$
                   & {\footnotesize $0|$-$2\EM5$}
                   &
                   &
\\
\hline
$\Delta \overline{q}$
              &--1 & $-0.074 \pm 0.017$
                   & {\footnotesize -$0.04|0$}
                   & $-0.072 \pm 0.015$
                   & {\footnotesize -$0.048|0$}
                   &
                   &
\\
              &  0 & $-0.003 \pm 0.001$
                   & {\footnotesize -$2\EM4|0$}
                   & $-0.002 \pm 5\EM4$
                   & {\footnotesize -$2\EM4|0$}
                   &
                   &
\\
              &  1 & $-4\EM4 \pm 1\EM4$
                   & {\footnotesize $0|0$}
                   & $-2\EM4 \pm 6\EM5$
                   & {\footnotesize $0|0$}
                   &
                   &
\\
              &  2 & $-8\EM5 \pm 2\EM5$
                   & {\footnotesize $0|0$}
                   & $-4\EM5 \pm 1\EM5$
                   & {\footnotesize $0|0$}
                   &
                   &
\\
\hline
$\Delta G$  &--1 & $ 1.026 \pm 0.549$
                   & {\footnotesize $0.04|1\EM5$}
                   & $0.931 \pm 0.669$
                   & {\footnotesize $0.191|0$}
                   &
                   &
\\
              &  0 & $0.184 \pm 0.103$
                   & {\footnotesize $5\EM4|1\EM5$}
                   & $0.100 \pm 0.075$
                   & {\footnotesize $0.002|0$}
                   &
                   &
\\
              &  1 & $0.050 \pm 0.028$
                   & {\footnotesize $1\EM5|1\EM5$}
                   & $0.022 \pm 0.017$
                   & {\footnotesize $2\EM5|0$}
                   &
                   &
\\
              &  2 & $0.017 \pm 0.010$
                   & {\footnotesize $0|1\EM5$}
                   & $0.006 \pm 0.005$
                   & {\footnotesize $0|0$}
                   &
                   &
\\
\hline
\hline
\end{tabular}
\end{center}
\normalsize

\vspace{2mm}
\noindent
{\sf Table~7: Moments of the NLO parton densities and their
combinations for the parameterizations {\tt ISET=3,4} at $Q^2 =
4~\GeV^2$. The value
of the respective moment integrating only outside the $x$--range in which
currently deep--inelastic scattering data are measured, $0.02 < x < 0.85$,
are given for comparison (lower$|$upper~part). The errors are the 
$1\sigma$ correlated errors derived
in the present analysis form the polarization asymmetry world data.
The values of corresponding lattice measurements, cf.~\cite{LHSE}, 
are shown for comparison. For $n=0,1$ for the values of the QCDSF
collaboration no continuum extrapolation was performed.}
\renewcommand{\arraystretch}{1}
%
\renewcommand{\arraystretch}{1.3}
\begin{center}
\begin{tabular}{|l|r|r|r|r|r|r|r|}
\hline \hline 
\multicolumn{1}{|c|}{Distribution}&
\multicolumn{1}{ c|}{{\tt ISET=3}}&
\multicolumn{1}{ c|}{{\tt ISET=4}}&
\multicolumn{1}{ c|}{ABFR~\cite{ABFR}}&
\multicolumn{1}{ c|}{GRSV~\cite{GRSV}}&
\multicolumn{1}{ c|}{AAC~\cite{AAC}}\\
\hline\hline
$\Delta u_v$  & $0.926 \pm 0.071$
              & $0.926 \pm 0.062$
&                & 0.9206      & 0.9278      \\
$\Delta d_v$  &    
$-0.341 \pm 0.123$
&$-0.341 \pm 0.103$
&                & --0.3409      & --0.3416
\\
$\Delta u                      $
& $0.851 \pm 0.075$
& $0.854 \pm 0.066$
& $\eta_u =~~0.692$  &  0.8593      &   0.8399   \\
$\Delta d                      $
& $-0.415 \pm 0.124$
& $-0.413 \pm 0.104$
& $\eta_d= -0.418$               & --0.4043      & --0.4295   \\
$\Delta \overline{q}$
&       $-0.074  \pm 0.017$ &   $-0.072 \pm 0.015$
& 
&--0.0625 &--0.0879    \\
$\Delta G  $  & $1.026  \pm 0.549$
&       $0.931 \pm 0.669$
&     $1.262$    & 0.6828    &  0.8076      \\
\hline\hline
\end{tabular}
\end{center}
\normalsize

\vspace{2mm}
\noindent
{\sf Table~8: Comparison of the first moments of the polarized parton
densities in NLO in the $\overline{\rm MS}$ scheme
at $Q^2= 4\GeV^2$ for different sets of recent parton
parameterizations. For the ABFR-analysis~\cite{ABFR} the values
$\eta_{u,d}$ are the first moments of $\Delta u + \Delta \overline{u}$
and
$\Delta d + \Delta \overline{d}$, respectively, and
$\Delta s + \Delta \overline{s} = -0.081$.
}
\renewcommand{\arraystretch}{1}
%

\newpage

\section{Figures}
\label{sec:fig}
\begin{figure}[htb]
\begin{center}
\includegraphics[angle=0, width=14.0cm]{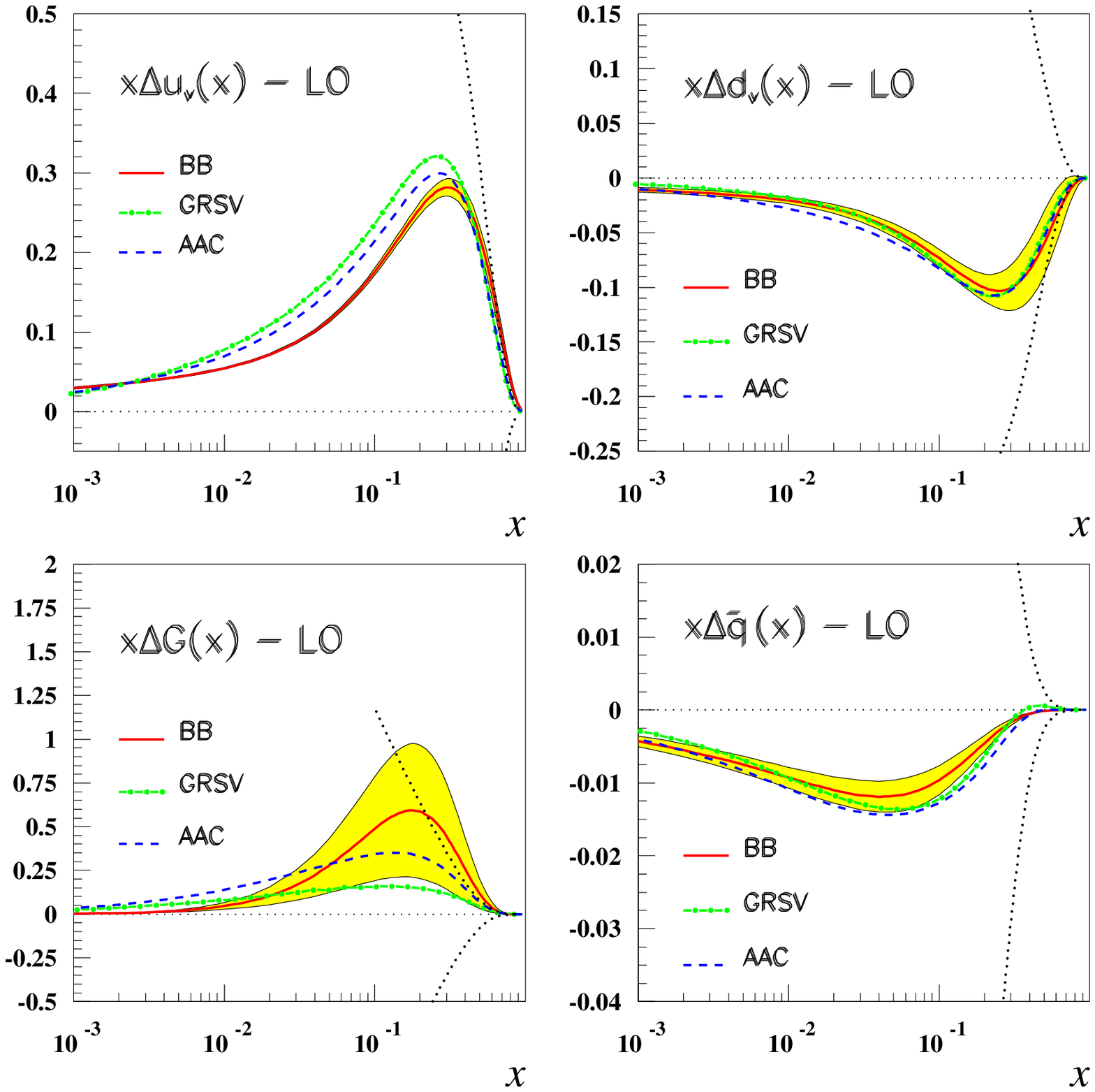} 
\end{center}
\caption{\label{dqlo}
LO polarized parton distributions at the input scale $Q_0^2 = 4.0~\GeV^2$,
{\tt ISET = 1},
(solid line) compared to results obtained by GRSV (dashed--dotted
line) \cite{GRSV} and AAC (dashed line) \cite{AAC}. The shaded areas
represent the fully correlated $1\sigma$ error bands calculated by 
Gaussian error propagation. The dark
dotted lines indicate the positivity bounds
choosing the distributions \cite{GRV} for reference.
}
\end{figure}
\newpage
\begin{figure}[tbp]
\begin{center}
\includegraphics[angle=0, width=14.0cm]{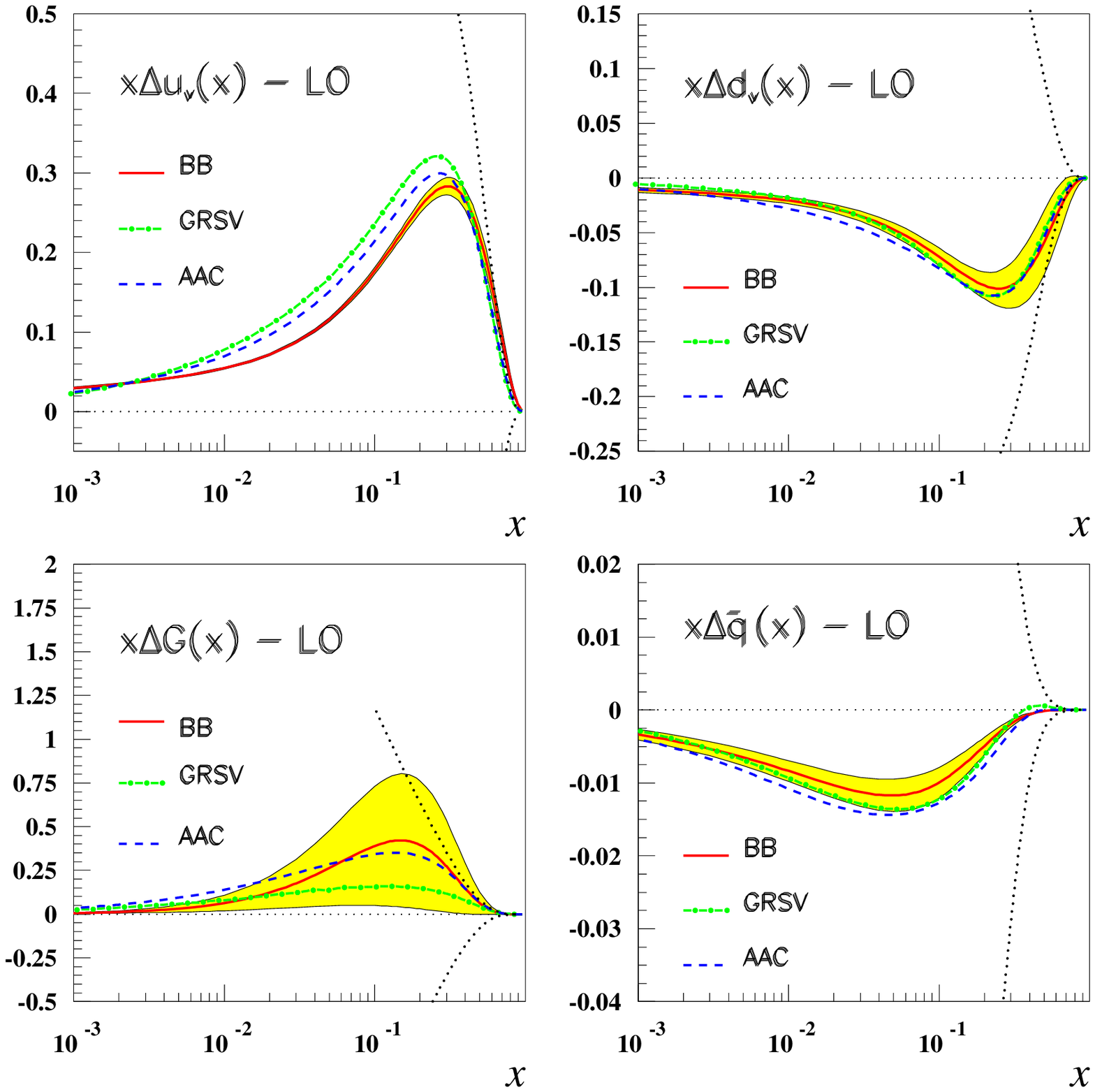}
\end{center}
\caption[xx]{\label{dqloB}
LO polarized parton distributions at the input scale $Q_0^2 = 4.0~\GeV^2$,
{\tt ISET=2},
(solid line) compared to results obtained by GRSV (dashed--dotted
line) \cite{GRSV} and AAC (dashed line) \cite{AAC}. The shaded areas
represent the fully correlated $1\sigma$ error bands calculated by 
Gaussian error propagation. The dark
dotted lines correspond to the positivity
bounds according to the parameterization~\cite{GRV}. }
\end{figure}
\newpage
\begin{figure}[htb]
\begin{center}
\includegraphics[angle=0, width=14.0cm]{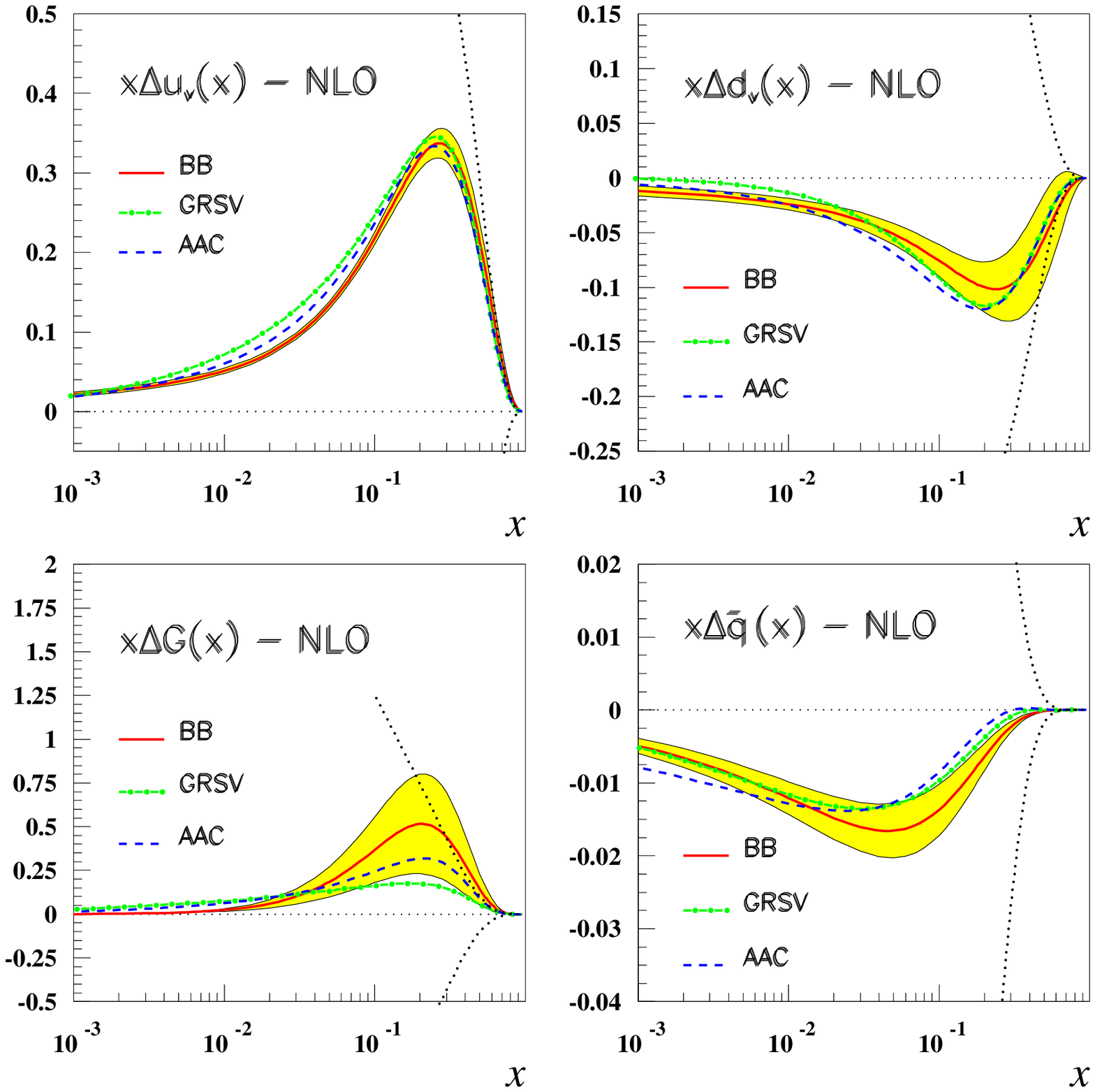}
\end{center}
\caption{\label{dqnlo}
NLO polarized parton distributions at the input scale 
$Q_0^2 = 4.0~\GeV^2$, {\tt ISET=3},
(solid line) compared to results obtained by GRSV (dashed--dotted
line) \cite{GRSV} and AAC (dashed line) \cite{AAC}. The shaded areas
represent the fully correlated $1\sigma$ error bands calculated by 
Gaussian error propagation. The dark
dotted lines correspond to the positivity
bounds choosing \cite{GRV} for reference.}
\end{figure}
\newpage
\begin{figure}[htb]
\begin{center}
\includegraphics[angle=0, width=14.0cm]{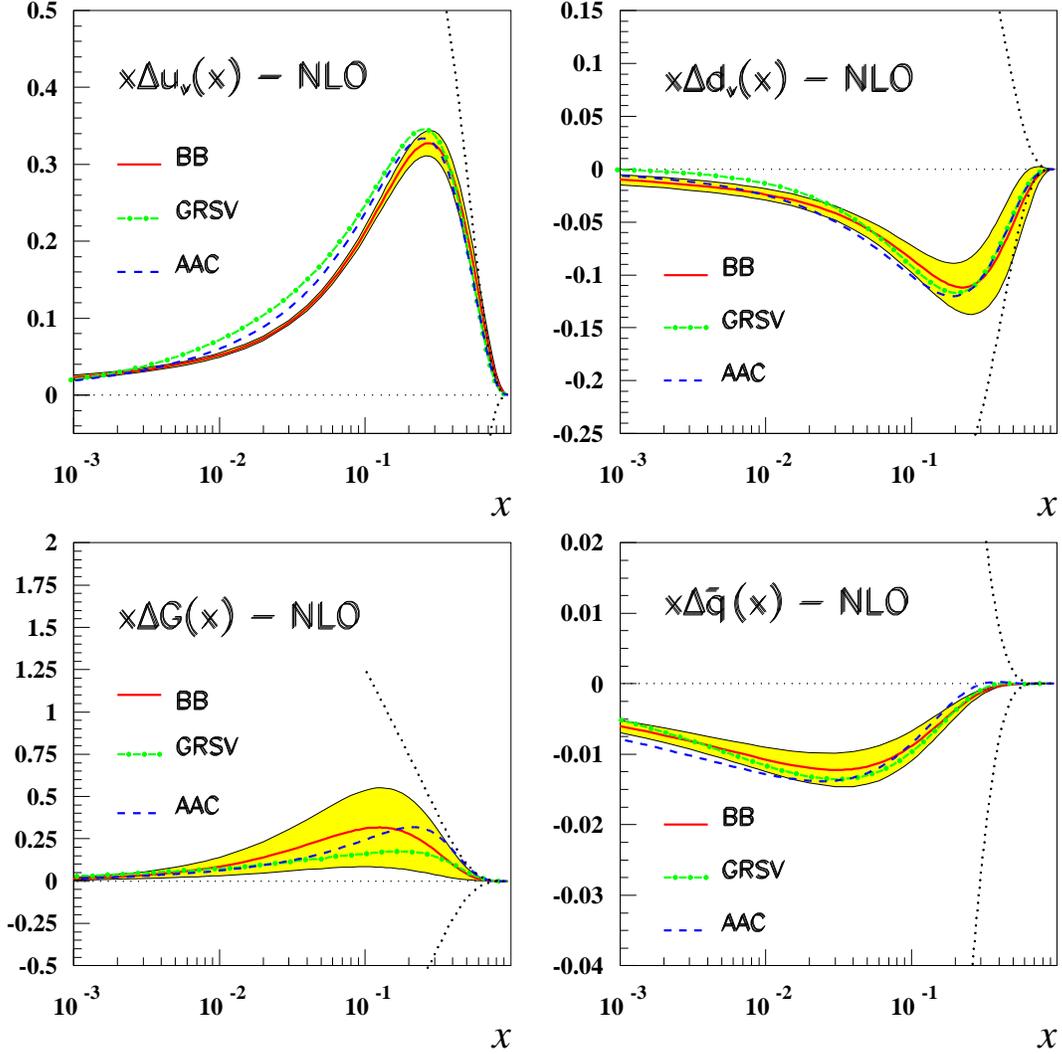}
\end{center}
\caption{\label{dqnloB}
NLO polarized parton distributions at the input scale 
$Q_0^2 = 4.0~\GeV^2$, {\tt ISET=4},
(solid line) compared to results obtained by GRSV (dashed--dotted
line) \cite{GRSV} and AAC (dashed line) \cite{AAC}. The shaded areas
represent the fully correlated $1\sigma$ error bands calculated by 
Gaussian error propagation. The dark
dotted lines indicate the positivity
bound if reference is taken to the distributions \cite{GRV}.}
\end{figure}
\newpage
\begin{figure}[htb]
\begin{center}
\includegraphics[angle=0, width=14.0cm]{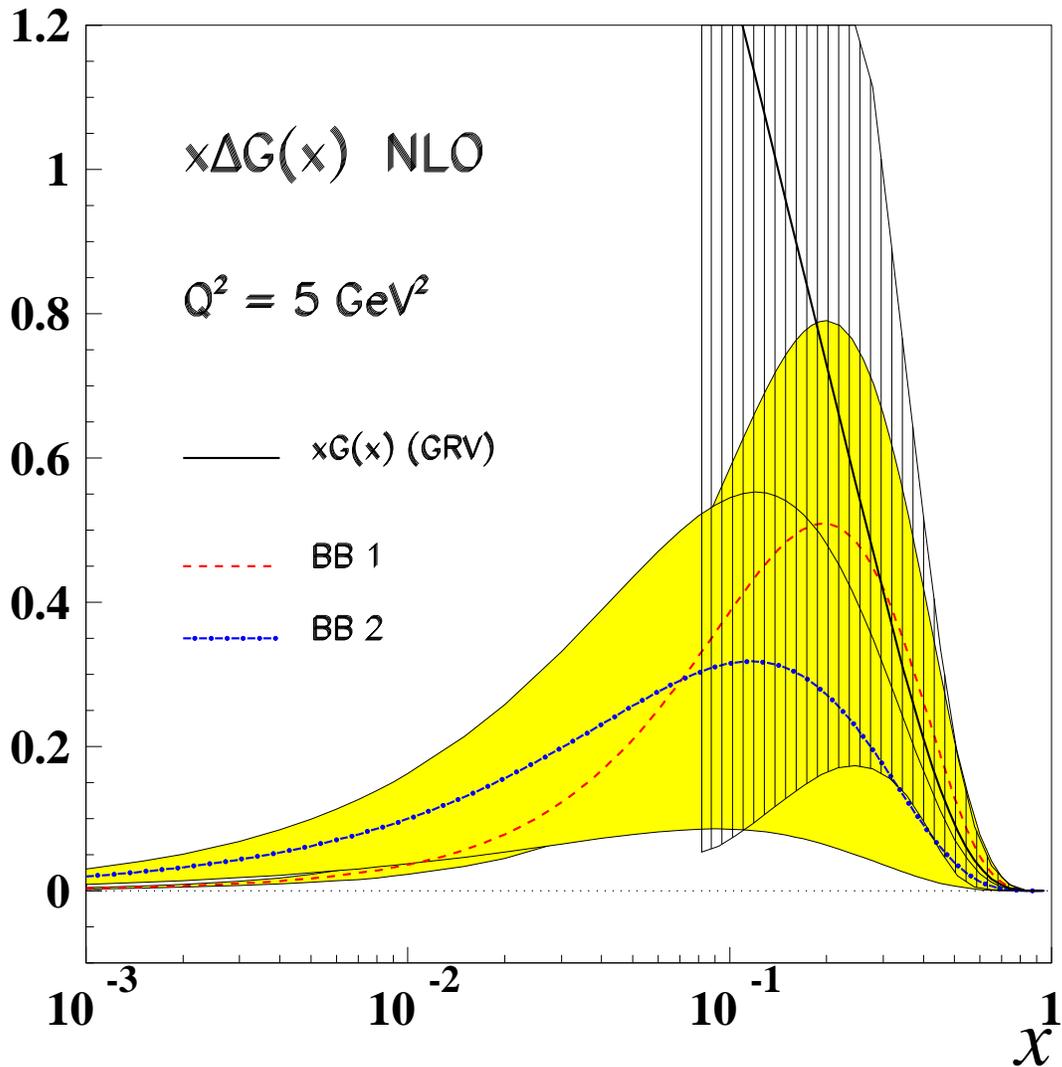}
\end{center}
\caption{\label{dqnloA}
NLO polarized momentum  distribution for the gluon  at the input scale 
$Q_0^2 = 5.0~\GeV^2$
for {\tt ISET=3} (BB1, dashed line) and {\tt ISET =4} 
(BB2, dash-dotted line)  with 1$\sigma$ error bands shown (shaded
areas). The solid line corresponds to the unpolarized distribution
\cite{GRV}. To the latter we added the experimental error of the
unpolarized gluon distribution as determined in the H1 
experiment~\cite{H1} (hatched area), see also [48]. 
}
\end{figure}
\newpage
\begin{figure}[htb]
\begin{center}
\includegraphics[angle=0, width=14.0cm]{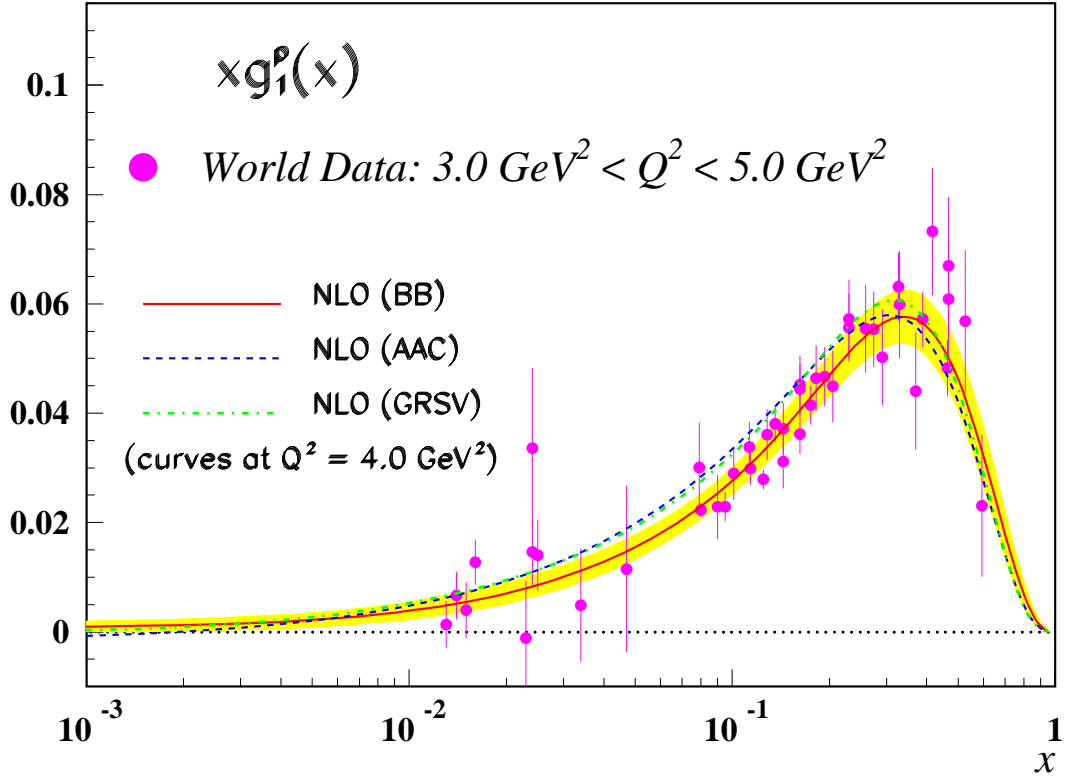}
\end{center}
\caption{\label{xg1p}
The structure function $xg_1^p$ measured in the interval $3.0~\GeV^2 <
Q^2 < 5.0~\GeV^2$ as function of $x$. The structure function
$g_1^p(x,Q^2)$ has been derived from the
world asymmetry data. The error bars shown are the statistical errors
only.
The distribution is well 
described by our QCD NLO curve {\tt ISET=3}
(solid line) at $Q^2 = 4.0~\GeV^2$ and
its fully correlated $1\sigma$ error band calculated by Gaussian error
propagation (shaded area). Also shown are the QCD NLO
curves at the same value of $Q^2$  obtained by AAC (dotted line) 
\cite{AAC} and  GRSV (dashed line) \cite{GRSV} for comparison.}
\end{figure}
\newpage
\begin{figure}[htb]
\begin{center}
\includegraphics[angle=0, width=14.0cm]{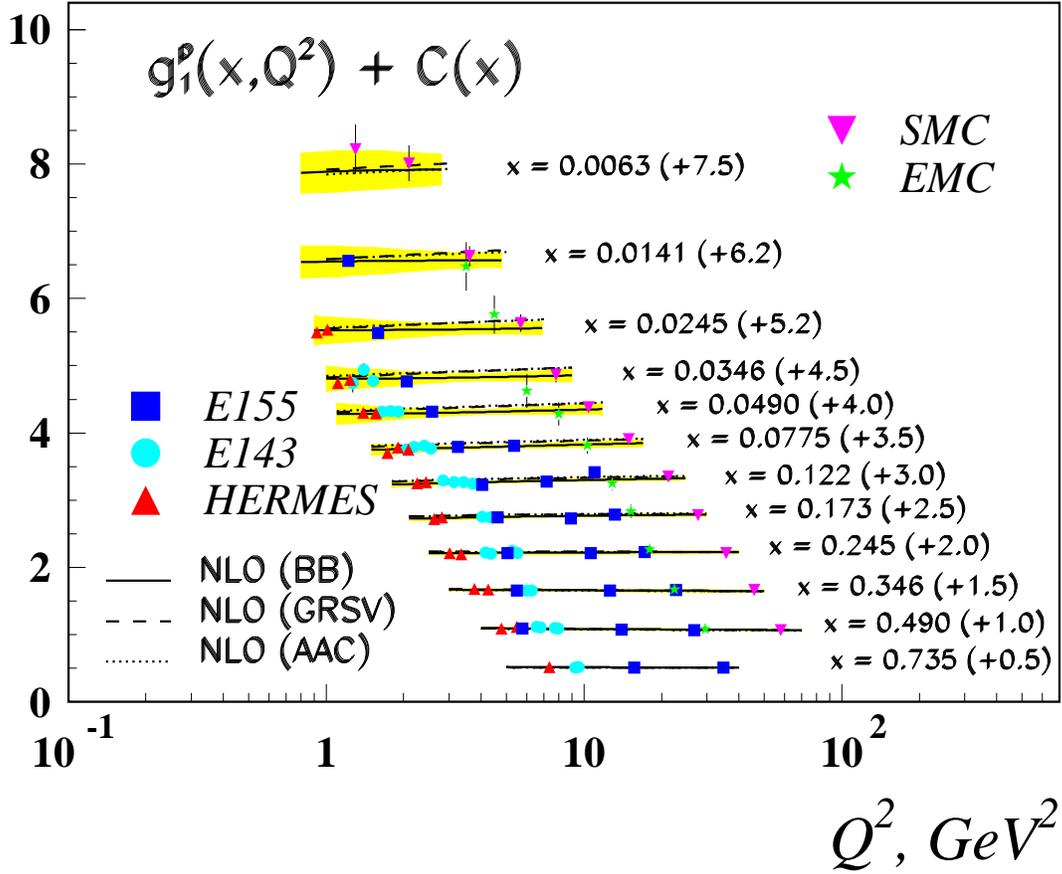}
\end{center}
\caption{\label{g1pvsq2}
The polarized structure function $g_1^p$ as function of $Q^2$ in
intervals of $x$. The error bars shown are the statistical and
systematic uncertainties added in quadrature. The data are well
described by our QCD NLO curves (solid lines), {\tt ISET=3},
and its fully correlated
$1\sigma$ error bands calculated by Gaussian error propagation (shaded
area). The values of $C(x)$ are given in parentheses.
Also shown  are the QCD NLO curves obtained by
AAC (dashed lines) \cite{AAC} and GRSV (dashed--dotted lines) 
\cite{GRSV} for comparison.}
\end{figure}
\newpage
\begin{figure}[htb]
\begin{center}
\includegraphics[angle=0, width=14.0cm]{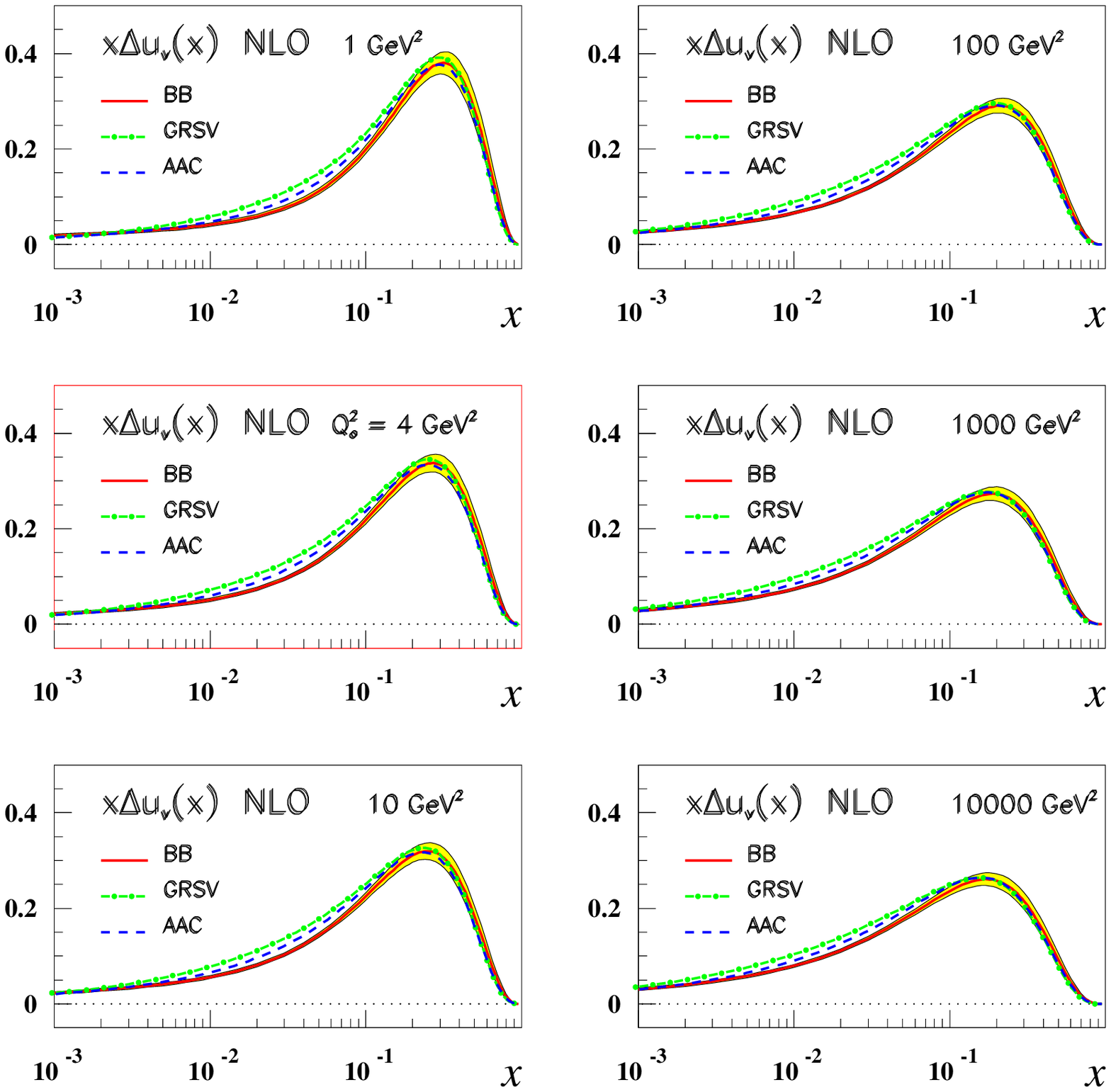}
\end{center}
\caption{\label{evol_uv}
The polarized parton distribution $x\Delta u_v$, {\tt ISET=3},
evolved up to values of
$Q^2 = 10,000~\GeV^2$ (solid lines) compared to results
obtained by GRSV (dashed--dotted lines) \cite{GRSV} and AAC (dashed
lines) \cite{AAC}. The shaded areas represent the fully correlated
$1\sigma$ error bands from our analysis calculated by Gaussian error
propagation.}
\end{figure}
\newpage
\begin{figure}[htb]
\begin{center}
\includegraphics[angle=0, width=14.0cm]{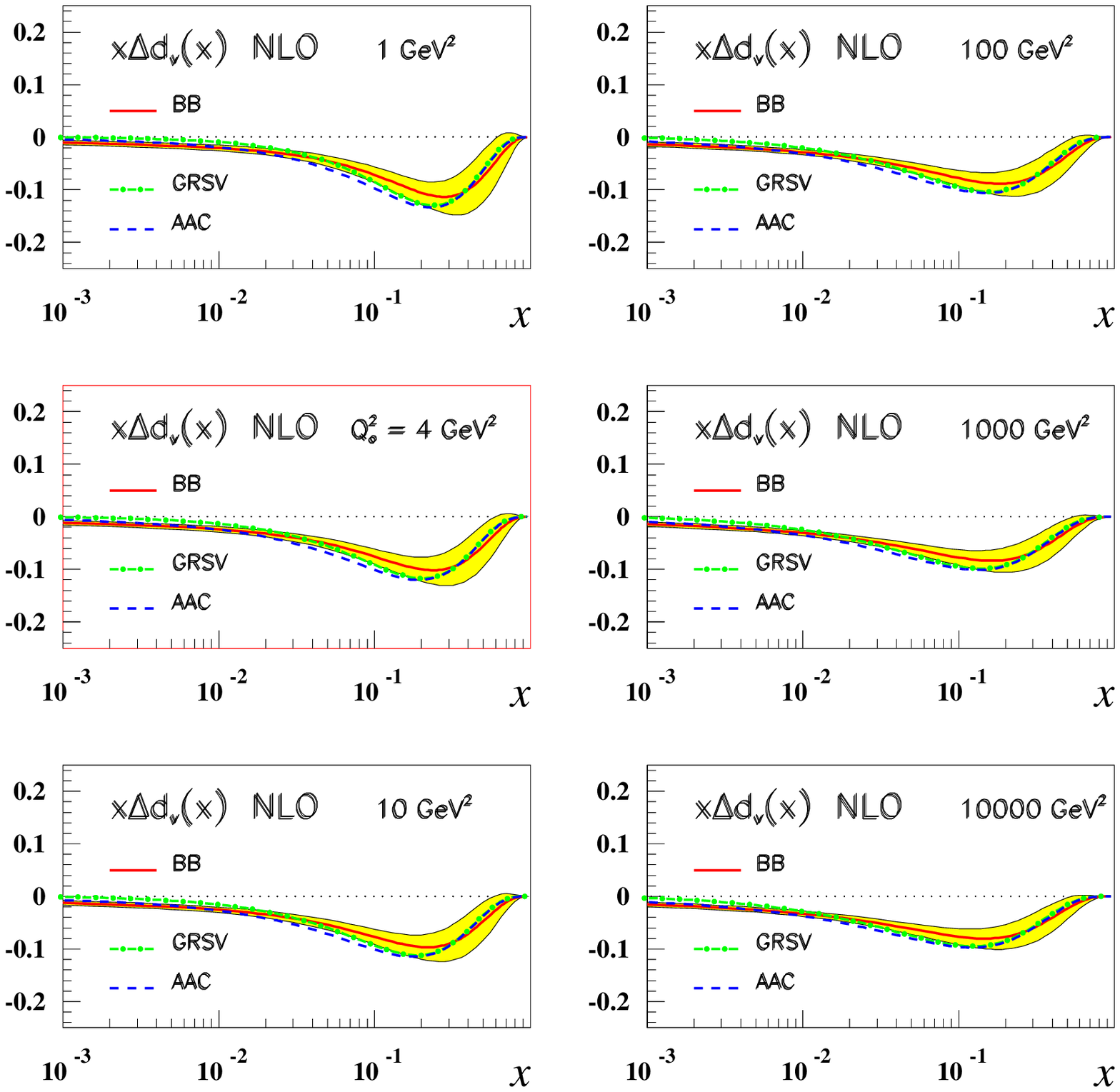}
\end{center}
\caption{\label{evol_dv}
The polarized parton distribution $x\Delta d_v$, {\tt ISET = 3},
evolved up to
values of $Q^2 = 10,000~\GeV^2$ (solid lines) compared to results
obtained by GRSV (dashed--dotted lines) \cite{GRSV} and AAC (dashed
lines) \cite{AAC}. The shaded areas represent the fully correlated
$1\sigma$ error bands from our analysis calculated by Gaussian error
propagation.}
\end{figure}
\newpage
\begin{figure}[htb]
\begin{center}
\includegraphics[angle=0, width=14.0cm]{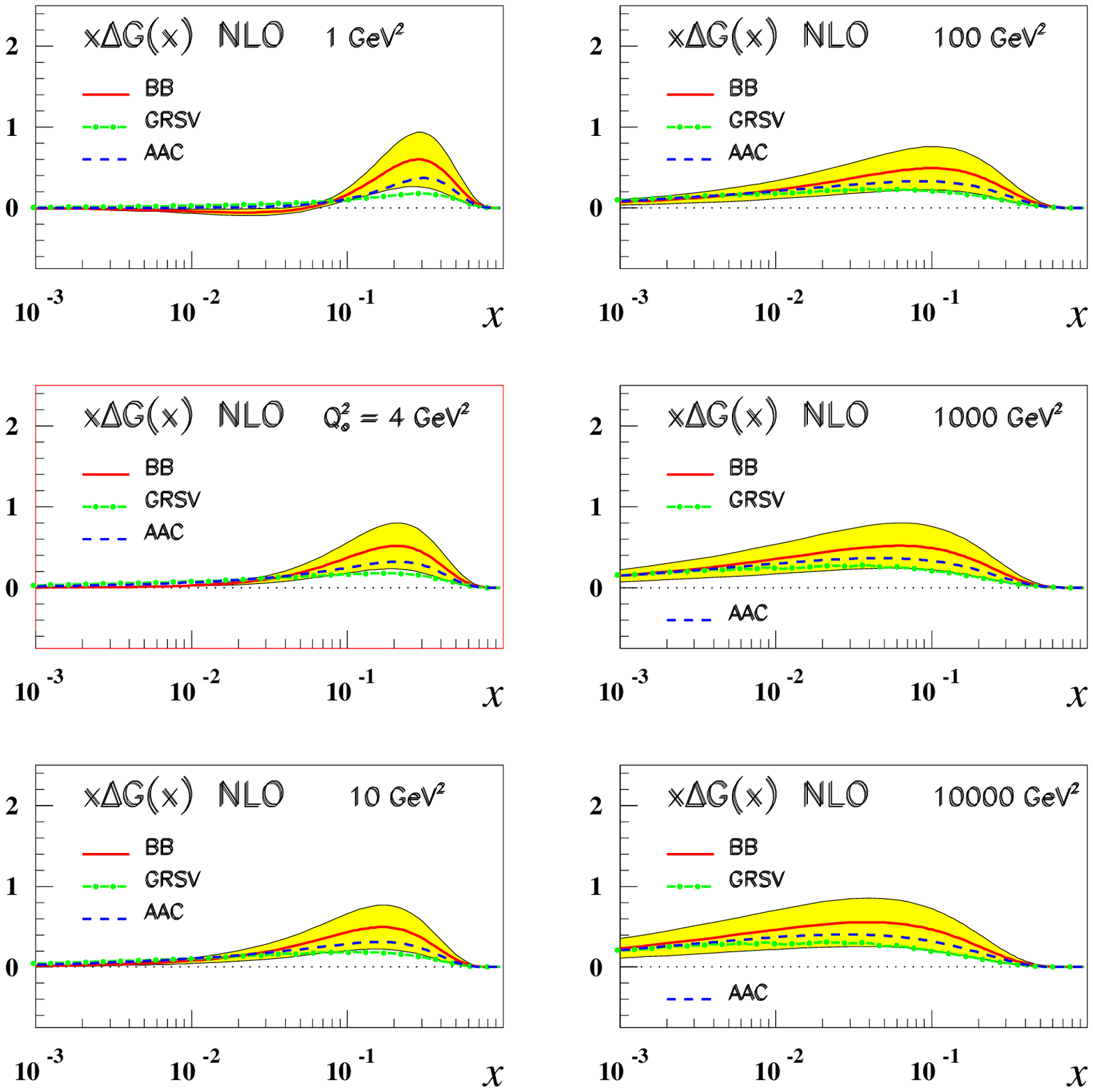}
\end{center}
\caption{\label{evol_gl}
The polarized parton distribution $x\Delta G$, {\tt ISET = 3},
evolved up to 
values of $Q^2 = 10,000~\GeV^2$ (solid lines) compared to results
obtained by GRSV (dashed--dotted lines) \cite{GRSV} and AAC (dashed
lines) \cite{AAC}. The shaded areas represent the fully correlated
$1\sigma$ error bands from our analysis calculated by Gaussian error
propagation.}
\end{figure}
\newpage
\begin{figure}[htb]
\begin{center}
\includegraphics[angle=0, width=14.0cm]{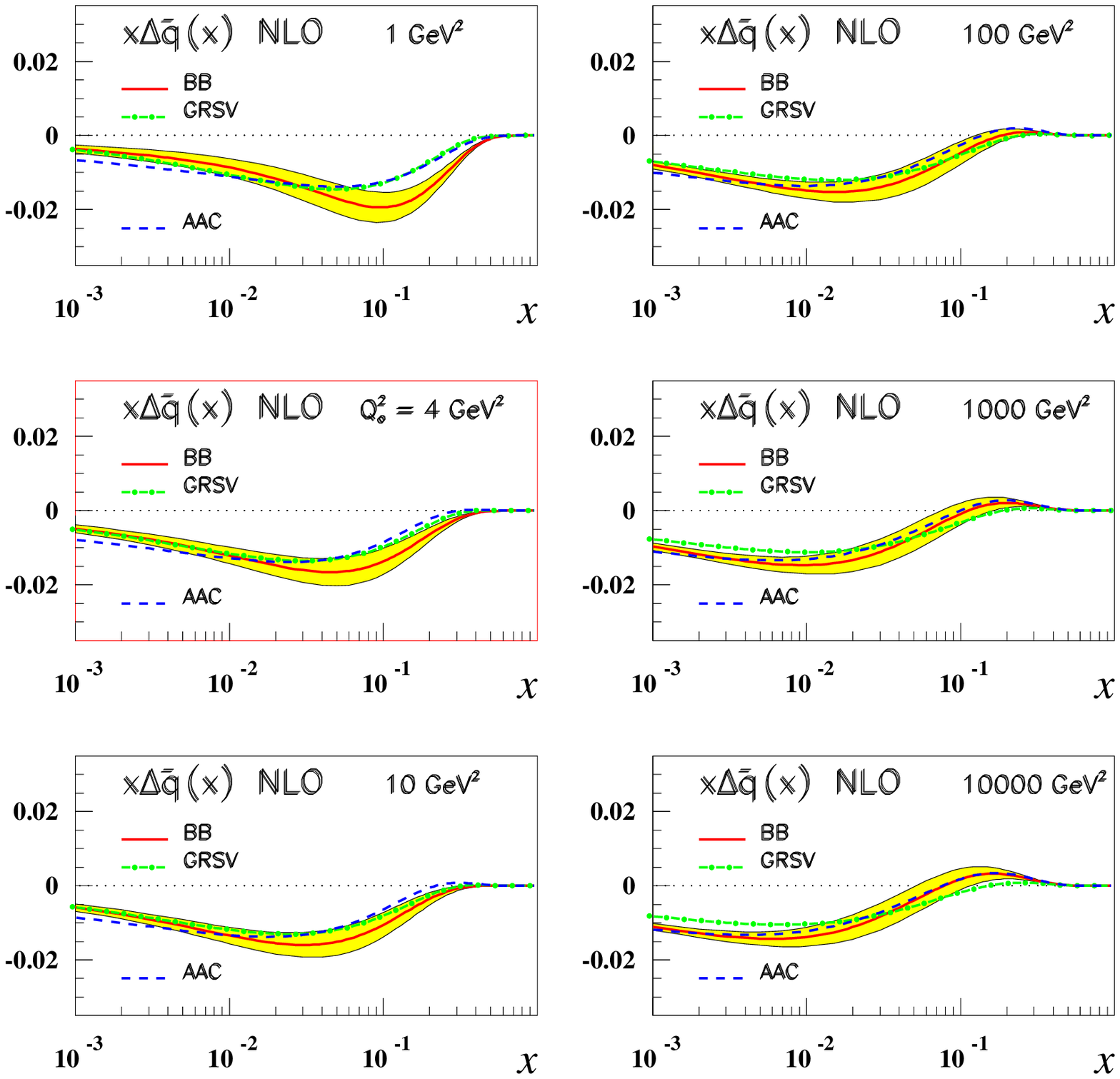}
\end{center}
\caption{\label{evol_qb}
The polarized parton distribution $x\Delta {\bar q}$, {\tt ISET =3},
evolved up to 
values of $Q^2 = 10,000~\GeV^2$ (solid lines) compared to results
obtained by GRSV (dashed--dotted lines) \cite{GRSV} and AAC (dashed
lines) \cite{AAC}. The shaded areas represent the fully correlated
$1\sigma$ error bands from our analysis calculated by Gaussian error
propagation.}
\end{figure}
\newpage
\begin{figure}[tbp]
\begin{center}
\includegraphics[angle=0, width=14.0cm]{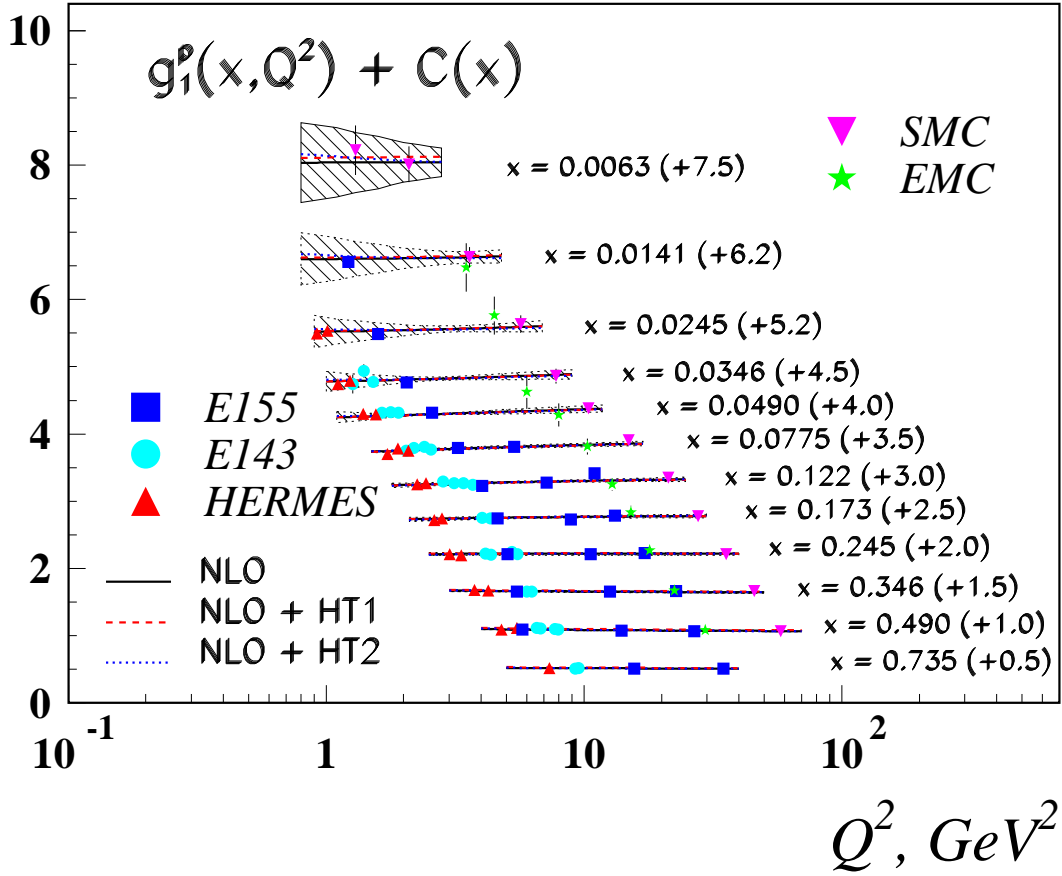}
\end{center}
\caption[xx]{Model fit to potential power corrections in $g_1(x,Q^2)$
as extracted from the world polarization asymmetry data in the present
analysis (see text). Dashed line: model I, Eq.~(\ref{eqHT1}); 
dotted line: model II, Eq.~(\ref{eqHT2}). 
The full lines correspond to the parameterization
({\tt ISET=4}) in the present analysis, to which the corresponding
power correction model induces a perturbation. The shaded area 
corresponds to the $1\sigma$ correlated error.}
\end{figure}
\newpage
\begin{figure}[htb]
\begin{center}
\includegraphics[angle=0, width=14.0cm]{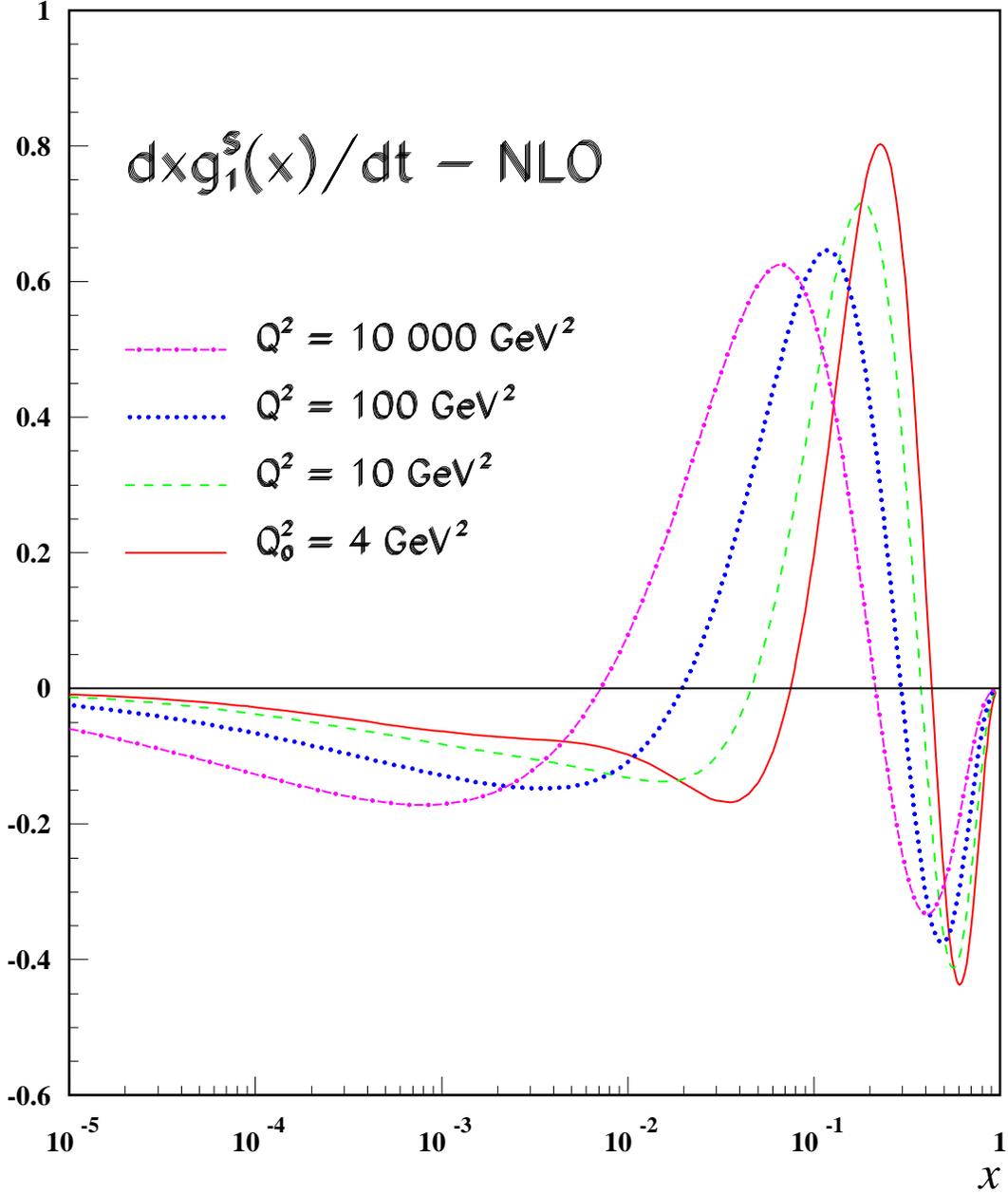}
\end{center}
\caption{\label{dg1dt} The evolution of $\partial xg_1^S(x,Q^2) /
\partial t$, the slope of the polarized structure function
$xg_1^p(x,Q^2)$ of the proton (singlet contribution) with respect to 
$t = -2 / \beta_0 \ln(\alpha_s(Q^2) / \alpha_s(Q_0^2))$ \cite{DLY}. The
slope was determined from a fit to  $g_1^p(x,Q^2)$, see text.}
\end{figure}
\newpage
\clearpage
\portrait

%
\end{document}